\RequirePackage{ifpdf}
\ifpdf 
\documentclass[pdftex]{sigma}
\else
\documentclass{sigma}
\fi

\def\l[{\left[}
\def\r]{\right]}

\newcommand{\ticup}{{\mbox{\footnotesize$\cup $}}}

\numberwithin{equation}{section}
\numberwithin{theorem}{section}
\numberwithin{proposition}{section}
\numberwithin{definition}{section}
\numberwithin{remark}{section}

\begin{document}

\allowdisplaybreaks

\renewcommand{\PaperNumber}{078}

\FirstPageHeading

\ShortArticleName{Deligne--Beilinson Cohomology and Abelian Link Invariants}

\ArticleName{Deligne--Beilinson Cohomology\\ and Abelian Link Invariants}

\Author{Enore GUADAGNINI~$^\dag$ and Frank THUILLIER~$^\ddag$}

\AuthorNameForHeading{E. Guadagnini and F. Thuillier}

\Address{$^\dag$~Dipartimento di Fisica ``E. Fermi'' dell'Universit\`a di Pisa\\
 \hphantom{$^\dag$}~and  Sezione di Pisa dell'INFN, Italy}
\EmailD{\href{mailto:enore.guadagnini@df.unipi.it}{enore.guadagnini@df.unipi.it}}
\URLaddressD{\url{http://www.df.unipi.it/~guada/}}

\Address{$^\ddag$~LAPTH, Chemin de Bellevue, BP 110, F-74941
Annecy-le-Vieux cedex, France}
\EmailD{\href{mailto:frank.thuilier@lapp.in2p3.fr}{frank.thuilier@lapp.in2p3.fr}}
\URLaddressD{\url{http://lappweb.in2p3.fr/~thuillie/}}

\ArticleDates{Received July 14, 2008, in f\/inal form October 27,
2008; Published online November 11, 2008}

\Abstract{For the Abelian Chern--Simons f\/ield  theory,  we  consider  the quantum functional integration over the Deligne--Beilinson cohomology classes  and we derive the main properties of the observables in a generic closed orientable 3-manifold.  We present an explicit path-integral non-perturbative computation of the Chern--Simons link invariants in the case of the torsion-free 3-manifolds $S^3$, $S^1 \times S^2$ and $S^1 \times \Sigma_g$.}


\Classification{81T70; 14F43; 57M27}


\section{Introduction}\label{sec1}

The topological quantum f\/ield theory which is def\/ined by the Chern--Simons action can be used to compute  invariants of links in  3-manifolds \cite{SW, HA, PO, W1}.  The algebraic structure of these invariants, which is based on the properties of the characters  of simple Lie groups, is rather general.  In fact, these invariants can also be def\/ined by means of skein relations  or of quantum group Hopf algebra methods \cite{JO, RT}.

In the standard quantum f\/ield theory approach, the gauge invariance group of  the Abelian Chern--Simons theory is given by the set of local $U(1)$ gauge transformations and the observables can directly be computed  by means of perturbation theory when the ambient space is ${\mathbb R}^3$ (the result also provides the values of the link invariants in $S^3$). For a  nontrivial 3-manifold  $M_3$, the standard gauge theory approach presents some technical dif\/f\/iculties, and one open problem of the quantum Chern--Simons theory is to produce directly the functional integration in the case of a generic 3-manifold $M_3$.  In this article we will show how this can be done, at least for a certain class of nontrivial 3-manifolds, by using the Deligne--Beilinson cohomology.
 We shall concentrate on the Abelian Chern--Simons invariants; hopefully, the method that we present will possibly admit an extension to the non-Abelian case.

The  Deligne--Beilinson approach presents some remarkable aspects.
The space of classical f\/ield conf\/igurations which are factorized out by gauge invariance is enlarged with respect to the standard f\/ield theory formalism. Indeed, assuming that the quantum amplitudes given by the exponential of the holonomies~-- which are associated with oriented loops~--- represent a~complete set of observables, the functional integration  must locally correspond  to a sum over 1-forms modulo forms with integer periods, i.e.\ it  must correspond to a sum over Deligne--Beilinson classes. In this new approach, the structure  of the functional space admits a natural description in terms of the homology groups of the 3-manifold $M_3$.  This structure will be used to compute the Chern--Simons observables, without the use of perturbation theory,   on a class of  torsion-free manifolds.

The article is organized as follows. Section~\ref{sec2} contains a description of the basic properties of the Deligne--Beilinson cohomology and of the distributional extension of the space of the equivalence classes.  The framing procedure is  introduced in Section~\ref{sec3}. The general properties of the Abelian Chern--Simons theory are discussed in Section~\ref{sec4}; in particular,  non-perturbative proofs of the colour periodicity, of the ambient isotopy invariance and of the satellite relations are given.  The solution of the Chern--Simons theory on $S^3$ is presented in Section~\ref{sec5}. The computations of the observables for the manifolds $S^1 \times S^2$ and $S^1 \times \Sigma_g$ are produced in Sections~\ref{sec6} and \ref{sec7}.  Section~\ref{sec8} contains a brief description of the surgery rules that can be used to derive the link invariants in a generic 3-manifold, and it is checked that the results obtained by means of the Deligne--Beilinson cohomology and by means of the surgery method coincide.  Finally, Section~\ref{sec9} contains the conclusions.

\section[Deligne-Beilinson cohomology]{Deligne--Beilinson cohomology}\label{sec2}

The applications  of the Deligne--Beilinson (DB) cohomolgy \cite{Del71,B85,EV88,J88,Bry93} -- and of its various equivalent versions such as the
 Cheeger--Simons Dif\/ferential Characters \cite{CS73,K73/74}
or Sparks~\cite{HLZ03}~-- in quantum physics has been discussed
by various authors \cite{AO00,AO03,Alv84,G87,W88,Fr92,W99,Zuc00,Hop02}.
For instance, geometric quantization is based on classes of $U(1)$-bundles with connections, which are exactly DB classes of degree one (see Section~8.3 of \cite{Wo92}); and the
Aharanov--Bohm ef\/fect also admits a natural description in terms of DB cohomology.

In this article, we shall consider the computation of the Abelian  link invariants of the Chern--Simons theory by means of the DB cohomology.  Let $L$ be  an oriented  (framed and coloured) link in the 3-manifold $M_3$; one is interested in the ambient isotopy invariant which is def\/ined by the path-integral expectation value
\begin{gather}
\left\langle {\exp\left\{ {2i\pi \int_L A } \right\}} \right\rangle _{k}
\equiv \frac{\int {DA  \exp\left\{ {2i\pi k\int_{M_3 } {A\wedge dA} }
\right\} \exp\left\{ {2i\pi \int_L A } \right\}} }{\int {DA \exp\left\{ {2i\pi
k\int_{M_3 } {A\wedge dA} } \right\}} }  ,
\label{1}
\end{gather}
where the parameter $k$ represents the dimensionless coupling constant of the f\/ield theory. In equation (\ref{1}),  the holonomy associated with the link $L$ is def\/ined in terms of a $U\left( 1 \right)$-connec\-tion~$A$ on~$M_3 $; this holonomy is closely related to the classes of $U(1)$-bundles with connections that represent  DB cohomology classes. The Chern--Simons lagrangian~$A\wedge dA$  can be understood as a~DB cohomology class from the Cheeger--Simons Dif\/ferential Characters point of view, and it can also be interpreted as a DB ``square'' of $A$ which is def\/ined, as we shall see, by means of the DB $*$-product.

To sum up,  the DB cohomology appears to be the natural  framework which should  be used in order to compute the Chern--Simons expectation values (\ref{1}).   As we shall see, this will imply the quantization of the coupling constant $k$  and  it will actually provide the  integration measure~$DA$ with  a nontrivial structure which is related to the homology of the manifold~$M_3$. It should be noted that the gauge invariance of the Chern--Simons action and of the observables is totally included into the DB
setting: working with DB classes means that we have already taken the quotient
by gauge transformations.

Although we won't describe DB cohomology in full details, we shall now present  a few pro\-per\-ties of the DB cohomology that will be useful for the non-perturbative computation of the observables (\ref{1}).

\subsection{General properties}\label{sec2.1}

Let $M$ be a smooth oriented compact manifold without boundary of f\/inite
dimension $n$. The Deligne cohomology group of $M$ of degree $q$, $H_D^q \left(
{M,{\mathbb Z}} \right)$, can be described as the central term of the following
exact sequence
\begin{gather}
\label{2}
0\buildrel \over \longrightarrow {\Omega ^q\left( M \right)} \mathord{\left/
{\vphantom {{\Omega ^q\left( M \right)} {\Omega _{\mathbb Z}^q \left( M
\right)}}} \right. \kern-\nulldelimiterspace} {\Omega _{\mathbb Z}^q \left( M
\right)}\buildrel \over \longrightarrow H_D^q \left( {M,{\mathbb Z}}
\right)\buildrel \over \longrightarrow H^{q+1}\left( {M,{\mathbb Z}}
\right)\buildrel \over \longrightarrow 0  ,
\end{gather}
where $\Omega ^q\left( M \right)$ is the space of smooth $q$-forms on $M$,
$\Omega _{\mathbb Z}^q \left( M \right)$ the space of smooth closed $q$-forms with
integral periods on $M$ and $H^{q+1}\left( {M,{\mathbb Z}} \right)$ is the
$\left( {q+1} \right)^{th}$ integral cohomology group of $M$. This last
space can be taken as simplicial, singular or Cech. There
is another exact sequence into which $H_D^q \left( {M,{\mathbb Z}} \right)$ can
be embedded, namely
\begin{gather}
\label{3}
0\buildrel \over \longrightarrow H^q\left( {M,{\mathbb R} \mathord{\left/
{\vphantom {{\mathbb R} {\mathbb Z}}} \right. \kern-\nulldelimiterspace} {\mathbb Z}}
\right)\buildrel \over \longrightarrow H_D^q \left( {M,{\mathbb Z}}
\right)\buildrel \over \longrightarrow \Omega _{\mathbb Z}^{q+1} \left( M
\right)\buildrel \over \longrightarrow 0  ,
\end{gather}
where $H^q\left( {M,{\mathbb R} / {\mathbb Z}} \right)$ is the ${\mathbb R}
/ {\mathbb Z}$-cohomology group of $M$
\cite{Bry93,HLZ03,BGST05}.

One can compute $H_D^q \left( {M,{\mathbb Z}} \right)$
by using a (hyper) cohomological resolution of a double complex of Cech--de Rham
type, as explained for instance in \cite{EV88,BGST05}. In this approach, $H_D^q \left( {M,{\mathbb Z}} \right)$ appears as the
set of equivalence classes of DB cocycles which are def\/ined by sequences
$(\omega ^{(0,q)},\omega ^{(1,q-1)}$, $\dots, \omega ^{(q,0)},
\omega^{(q+1,-1)})$,
where $\omega ^{(p,q-p)}$ denotes a collection of smooth $(q-p)$-forms in the
intersections of degree $p$ of some open sets of a good open covering of $M$, and $\omega^{(q+1,-1)}$
is an \emph{integer} Cech $(p+1)$-cocyle for this open good covering of~$M$. These
forms satisfy cohomological descent equations of the type $\delta \omega ^{(p-1,q-p+1)} + d\omega
^{(p,q-p)} = 0$,
and the equivalence relation is def\/ined via the $\delta $ and $d$ operations, which are respectively the Cech and de Rham dif\/ferentials. The Cech--de~Rham point of view has the advantage of producing ``explicit'' expressions for representatives of DB classes in some good open covering of $M$.

\begin{definition}
\label{Definition2.1.1} Let $\omega $  be a $q$-form which is globally def\/ined on the manifold $M$.
We shall denote by $ [ \omega ] \in H_D^q \left( {M,{\mathbb Z}} \right)$ the DB class which, in the  Cech--de~Rham double complex approach, is  represented by the sequence  $(\omega ^{(0,q)}= \omega , \omega ^{(1,q-1)} =0 ,\dots ,\omega ^{(q,0)} =0 ,
\omega^{(q+1,-1)}=0)$.
\end{definition}

 From sequence (\ref{2}) it follows that $H_D^q \left( {M,{\mathbb Z}} \right)$ can be understood as an af\/f\/ine bundle over $H^{q+1}\left( {M,{\mathbb Z}} \right)$, whose f\/ibres have a typical underlying (inf\/inite dimensional) vector space
structure given by ${\Omega ^q\left( M \right)} / {\Omega _{\mathbb Z}^q \left( M \right)}$. Equivalently,
${\Omega ^q\left( M \right)} / {\Omega _{\mathbb Z}^q \left( M \right)}$
canonically acts on the f\/ibres of the bundle $H_D^q \left( {M,{\mathbb Z}}
\right)$ by translation.  From a geometrical point of view, $H_D^1 \left( {M,{\mathbb Z}}
\right)$ is canonically isomorphic to the space of equivalence classes of
$U\left( 1 \right)$-principal bundles with connections over~$M$ (see for
instance \cite{HLZ03,BGST05}). A generalisation of this idea has been proposed by means of
Abelian Gerbes (see for instance \cite{Bry93,MP00}) and Abelian Gerbes with connections over $M$.
In this framework, $H^{q+1}\left( {M,{\mathbb Z}} \right)$ classif\/ies equivalence classes of some Abelian Gerbes over $M$,
in the same way as $H^2\left( {M,{\mathbb Z}} \right)$ is the space which classif\/ies the $U\left( 1 \right)$-principal
bundles over $M$, and $H_D^q \left( {M,{\mathbb Z}} \right)$ appears as the set of equivalence classes of some Abelian Gerbes
with connections. Finally, the space $\Omega _{\mathbb Z}^q \left( {M} \right)$ can be interpreted as the group
of generalised  Abelian gauge transformations.

We shall  mostly be concerned with the cases $q=1$ and $q=3$. As
for $M$, we will consider the three dimensional cases $M_3 =S^3$, $M_3
=S^1\times S^2$ and $M_3 =S^1\times \Sigma _g $, where $\Sigma _g $ is a
Riemann surface of genus $g\ge 1$. In particular, $M_3 $ is oriented and
torsion free. In all these cases, the exact sequence (\ref{2}) for $q=3$ reads
\begin{gather*}
0\buildrel \over \longrightarrow {\Omega ^3\left( {M_3 } \right)}
/ {\Omega
_{\mathbb Z}^3 \left( {M_3 } \right)}\buildrel \over \longrightarrow H_D^3
\left( {M_3 ,{\mathbb Z}} \right)\buildrel \over \longrightarrow H^4\left( {M_3
,{\mathbb Z}} \right)=0\buildrel \over \longrightarrow 0 ,
\end{gather*}
where the f\/irst non trivial term reduces to
\begin{gather}
\label{5}
\frac{\Omega ^3\left( {M_3 } \right)}{\Omega _{\mathbb Z}^3 \left( {M_3 }
\right)}\cong \frac{{\mathbb R}}{{\mathbb Z}}  .
\end{gather}
The validity of equation (\ref{5}) can  easily be checked by using a volume form on $M_3 $.  By def\/inition, for any $\left( {\rho ,\tau _{\mathbb Z} } \right)\in \Omega ^3\left( {M_3 } \right)\times \Omega _{\mathbb Z}^3 \left( {M_3 } \right)$ one has
\begin{gather*}
\left[ {\rho +\tau _{\mathbb Z} } \right]=\left[ \rho \right] \in H_D^3
\left( {M_3 ,{\mathbb Z}} \right);
\end{gather*}
consequently
\begin{gather*}
H_D^3 \left( {M_3 ,{\mathbb Z}} \right)\simeq \frac{\Omega ^3\left( {M_3 }
\right)}{\Omega _{\mathbb Z}^3 \left( {M_3 } \right)}\cong \frac{{\mathbb R}}{{\mathbb Z}} .
\end{gather*}
These results imply that any Abelian 2-Gerbes on $M_3 $ is trivial
($H^4\left( {M_3 ,{\mathbb Z}} \right)=0)$, and the set of classes of Abelian
2-Gerbes with connections on $M_3 $ is isomorphic to ${\mathbb R}
/ {\mathbb Z}$.   In the less trivial case $q=1$,
sequence (\ref{2}) reads
\begin{gather}
\label{8}
0\buildrel \over \longrightarrow {\Omega ^1\left( {M_3 } \right)}
/ {\Omega
_{\mathbb Z}^1 \left( {M_3 } \right)}\buildrel \over \longrightarrow H_D^1
\left( {M_3 ,{\mathbb Z}} \right)\buildrel \over \longrightarrow H^2\left( {M_3
,{\mathbb Z}} \right)\buildrel \over \longrightarrow 0  .
\end{gather}

\begin{figure}[t]
\centerline{\includegraphics[width=60mm]{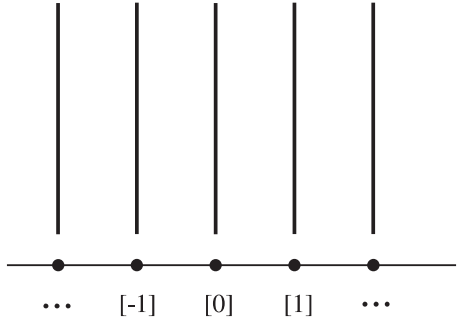}}
\caption{Presentation of the Deligne--Beilinson af\/f\/ine bundle
$H_D^1 \left( {S^1\times S^2,{\mathbb Z}} \right)$.}
\label{fig1}
\end{figure}

\noindent Still by def\/inition, for any $\left( {\eta ,\omega _{\mathbb Z} } \right)\in
\Omega ^1\left( {M_3 } \right)\times \Omega _{\mathbb Z}^1 \left( {M_3 }  \right)$ one has
\begin{gather*}
\left[ {\eta +\omega _{\mathbb Z} } \right]=\left[ \eta \right] \in H_D^1
\left( {M_3 ,{\mathbb Z}} \right )   .
\end{gather*}
When $H^2\left( {M_3 ,{\mathbb Z}} \right)=0$, sequence (\ref{8}) turns into a short
exact sequence; this also implies $H^1\left( {M_3 ,{\mathbb Z}} \right)=0$ due
to Poincar\'{e} duality.   For the 3-sphere $S^3$, the base space of $H_D^1 \left( {S^3,{\mathbb Z}} \right)$ is trivial.  Whereas, the bundle $H_D^1 \left( {S^1\times S^2,{\mathbb Z}} \right)$ has base space $H^2\left( {S^1\times S^2,{\mathbb Z}}
\right) \cong {\mathbb Z}$ and, as depicted in Fig.~\ref{fig1},
its f\/ibres are (inf\/inite dimensional) af\/f\/ine spaces whose underlying linear space identif\/ies with the quotient space
 ${\Omega ^1\left({S^1\times S^2} \right)} / {\Omega _{\mathbb Z}^1 \left(
{S^1\times S^2} \right)}$.  In the general case $M_3 =S^1\times \Sigma _g $ with $ g \geq 1 $,  the base space  $H^2\left( {S^1\times \Sigma _g ,{\mathbb Z}} \right)$ is isomorphic to ${\mathbb Z}^{2g+1}$.

Finally, one should note that sequence (\ref{8}) also gives  information on $\Omega _{\mathbb Z}^1 \left( {M_3 } \right)$ since its structure is mainly given by the  $H^1_D\left( {M_3 ,{\mathbb Z}}
\right)$. For instance, $\Omega _{\mathbb Z}^1 \left( {S^3} \right)=d\Omega
^0\left( {S^3} \right)$, all other cases being not so trivial.

\subsection{Holonomy and pairing}\label{sec2.2}

As we have already mentioned, DB cohomology is the natural
framework in which integration (or holonomy) of a $U\left( 1
\right)$-connection over 1-cycles of $M_3 $ can be def\/ined and generalised
to objects of higher dimension ($n$-connections and $n$-cycles). In fact
integration of a DB cohomology class $\left[ \chi \right]\in H_D^q \left(
{M,{\mathbb Z}} \right)$ over a $q$-cycle of $M$, denoted by  $C\in Z_q \left( M \right)$,
appears as a ${\mathbb R} / {\mathbb Z}$-valued linear pairing
\begin{gather}
 \left\langle {\;,\;} \right\rangle _q :  \ H_D^q \left( {M,{\mathbb Z}}
\right) \times Z_q \left( M \right)  \longrightarrow  {\mathbb R}
/ {{\mathbb Z}=S^1}, \nonumber\\
\phantom{\left\langle {\;,\;} \right\rangle _q :} \  \left( {\left[ \chi \right] , C} \right) \longrightarrow
\left\langle {\left[ \chi \right], C } \right\rangle _q \equiv \int\limits_C
{\left[ \chi \right]}  ,\label{11}
\end{gather}
which establishes the equivalence between DB cohomology and Cheeger--Simons characters \cite{CS73,K73/74,Bry93,HLZ03,BGST05}. Accordingly, a quantity such as
\begin{gather*}
\exp \left\{ {2i\pi \int_C {\left[ \chi \right]} } \right\}
\end{gather*}
is well def\/ined and corresponds to the fundamental
representation of ${\mathbb R} / {{\mathbb Z} = S^1} \simeq U\left(1
\right)$. Using the Chech--de Rham description of DB cocycles, one can then produce explicit formulae \cite{BGST05} for the pairing (\ref{11}).

Alternatively, (\ref{11}) can be seen as a dualising equation. More precisely, any $ C \in Z_q \left( M \right)$ belongs to the Pontriagin dual of $H_D^q \left( {M,{\mathbb Z}} \right)$, usually denoted by ${\rm Hom}\left( {H_D^q \left( {M,{\mathbb Z}} \right),S^1} \right)$, the pairing (\ref{11}) providing a canonical injection
\begin{gather}
\label{13}
Z_q \left( M \right)  \,\vec {\subset }\,  {\rm Hom}\left( {H_D^q \left( {M,{\mathbb Z}}
\right),S^1} \right)  .
\end{gather}
A universal result \cite{G58} about the Hom functor implies the validity of the exact sequences,
duali\-sing (\ref{2}) (via (\ref{3})),
\begin{gather}
\label{14}
0\buildrel \over \longrightarrow {\rm Hom}\big( {\Omega _{\mathbb Z}^{q+1} \left( M
\right),S^1} \big)\buildrel \over \longrightarrow {\rm Hom}\left( {H_D^q \left(
{M,{\mathbb Z}} \right),S^1} \right)\buildrel \over \longrightarrow
H^{n-q}\left( {M,{\mathbb Z}} \right)\buildrel \over \longrightarrow 0,
\end{gather}
where $H^{n-q}\left( {M,{\mathbb Z}} \right)\cong {\rm Hom}\left(
{H^q\left( {M,{\mathbb R} / {\mathbb Z}} \right),S^1} \right)$.

The space ${\rm Hom}\left( {H_D^q \left( {M,{\mathbb Z}} \right),S^1}
\right)$ also contains $H_D^{n-q-1} \left( {M,{\mathbb Z}} \right)$, so that $Z_q \left( M
\right)$ (or rather its canonical injection (\ref{13})) can be seen as lying on
the boundary of $H_D^{n-q-1} \left( {M,{\mathbb Z}} \right)$ (see details in~\cite{HLZ03}). Accordingly
\begin{gather}
\label{15}
Z_q \left( M \right)\oplus H_D^{n-q-1} \left( {M,{\mathbb Z}} \right)\subset
{\rm Hom}\left( {H_D^q \left( {M,{\mathbb Z}} \right),S^1} \right),
\end{gather}
with the obvious abuse in the notation. Let us point out that, as suggested by  equation (\ref{15}),  one could represent integral cycles  by  currents which are singular (i.e.\ distributional) forms. This issue will be discussed in detail in next subsection.

Now, sequence (\ref{14}) shows  that ${\rm Hom}\left( {H_D^q \left( {M,{\mathbb Z}} \right),S^1} \right)$ is also an af\/f\/ine bundle with base space $H^{n-q}\left( {M,{\mathbb Z}} \right)$. In particular, let us consider the case in which $n=3$ and $q=1$; on the one hand,
 Poincar\'{e} duality implies
\begin{gather*}
H^{n-q}\left( {M,{\mathbb Z}} \right)=H^2\left( {M_3 ,{\mathbb Z}} \right)\cong
H^1\left( {M_3 ,{\mathbb Z}} \right) .
\end{gather*}
On the other hand, one has
\begin{gather*}
H_D^1 \left( {M,{\mathbb Z}} \right)\subset {\rm Hom}\left( {H_D^1 \left( {M,{\mathbb Z}}
\right),S^1} \right)  ,
\end{gather*}
 and, because of the Pontriagin duality,
\begin{gather*}
Z_1 \left( M \right)\oplus H_D^1 \left( {M,{\mathbb Z}} \right)\subset {\rm Hom}\left(
{H_D^1 \left( {M,{\mathbb Z}} \right),S^1} \right)  .
\end{gather*}
This is somehow reminiscent of  the
self-dual situation in the case of four dimensional manifolds and curvature.

\subsection{The product}\label{sec2.3}

The pairing (\ref{11}) is actually related to
another pairing of DB cohomology groups
\begin{gather}
\label{19}
H_D^p \left( {M,{\mathbb Z}} \right) \times H_D^q \left( {M,{\mathbb Z}}
\right)\longrightarrow H_D^{p+q+1} \left( {M,{\mathbb Z}}
\right) ,
\end{gather}
whose explicit description can be found for instance in
\cite{CS73,HLZ03,BGST05}. This pairing is known as the DB product (or DB $*$-product). It will be denoted by $\ast $.
In the Cech--de Rham approach, the DB product of the DB cocyle $(\omega ^{(0,p)},\omega ^{(1,p-1)},\dots ,\omega ^{(p,0)},
\omega^{(p+1,-1)})$ with the DB cocycle $(\eta ^{(0,q)},\eta ^{(1,q-1)},\dots ,\eta ^{(q,0)},
\eta^{(q+1,-1)})$ reads
\begin{gather}
\label{20bis}
\big( \omega^{(0,p)}\ticup d\eta^{(0,q)}, \ldots ,
\omega^{(p,0)}\ticup d\eta^{(0,q)} , b \omega^{(p+1,-1)} \ticup \eta^{(0,q)}, \ldots  ,   \omega^{(p+1,-1)}\ticup \eta^{(n-p,-1)} \big),
\end{gather}
where the product $\ticup$ is precisely def\/ined in \cite{BT82,EV88,BGST05}, for instance.

\begin{definition}
 \label{Definition2.3.1} Let us consider the sequence $(\eta ^{(0,q)},\eta ^{(1,q-1)},\dots ,\eta ^{(q,0)}, \eta^{(q+1,-1)})$, in which the components $\eta ^{(k-q,k)}$ satisfy the same descent equations as the components of a DB cocycle but, instead of smooth forms, these components are currents (i.e. distributional forms).  This allows to extend the (smooth) cohomology group $H_D^q \left( {M,{\mathbb Z}} \right)$ to a larger cohomology group that we will denote ${\widetilde H}_D^{q} \left( {M,{\mathbb Z}} \right)$.
\end{definition}

 Obviously, the DB product (\ref{20bis}) of a smooth DB cocycle with a distributional one is still well-def\/ined, and thus the pairing (\ref{19})  extends to
\begin{gather*}
H_D^p \left( {M,{\mathbb Z}} \right) \times {\widetilde H}_D^{ q} \left( {M,{\mathbb Z}}
\right) \longrightarrow {\widetilde H}_D^{ p+q+1} \left( {M,{\mathbb Z}}
\right)  .
\end{gather*}
Then, it can be checked \cite{BGST05} that any class $[\eta] \in {\widetilde H}_D^{ n-q-1} \left( {M,{\mathbb Z}} \right)$ canonically def\/ines a ${\mathbb R} / {\mathbb Z}$-valued linear pairing as in (\ref{11}) so that
\begin{gather*}
{\widetilde H}_D^{ n-q-1} \left( {M,{\mathbb Z}} \right) \subset {\rm Hom}\left( {H_D^q \left( {M,{\mathbb Z}} \right),S^1} \right).
\end{gather*}

It is important to note that, as it was shown in \cite{BGST05},  to any $C \in Z_q (M)$ there corresponds a~canonical DB class
$\left[ {\eta _C } \right]\in {\widetilde H}_D^{ n-q-1} \left( {M,{\mathbb Z}} \right)$ such that
\begin{gather*}
\exp \left\{ {2i\pi \int_C {\left[ \chi \right]} } \right\}=\exp \left\{
{2i\pi \int_M {\left[ \chi \right]\ast \left[ {\eta _C } \right]} }
\right\}  ,
\end{gather*}
for any $\left[ \chi \right] \in H_D^q \left( {M,{\mathbb Z}} \right)$. This means that we have the following sequence of canonical inclusions
\begin{gather*}
Z_q (M) \subset {\widetilde H}_D^{ n-q-1} \left( {M,{\mathbb Z}} \right) \subset
{\rm Hom}\left( {H_D^q \left( {M,{\mathbb Z}} \right),S^1} \right).
\end{gather*}
Let us point out the trivial inclusion
\begin{gather*}
H_D^{ n-q-1} \left( {M,{\mathbb Z}} \right) \subset {\widetilde H}_D^{ n-q-1} \left( {M,{\mathbb Z}} \right).
\end{gather*}

In the 3 dimensional case, let us consider the DB product
\begin{gather}
\label{22}
H_D^1 \left( {M_3 ,{\mathbb Z}} \right)\times H_D^1 \left( {M_3 ,{\mathbb Z}}
\right) \longrightarrow H_D^3 \left( {M_3 ,{\mathbb Z}}
\right)\cong {\mathbb R} \mathord{\left/ {\vphantom {{\mathbb R} {\mathbb Z}}} \right.
\kern-\nulldelimiterspace} {\mathbb Z}   .
\end{gather}
Starting from equation (\ref{22}) and extending it to
\begin{gather*}
H_D^1 \left( {M_3 ,{\mathbb Z}} \right)\times {\widetilde H}_D^{ 1} \left( {M_3 ,{\mathbb Z}}
\right) \longrightarrow {\widetilde H}_D^{ 3} \left( {M_3 ,{\mathbb Z}}
\right)\cong {\mathbb R} \mathord{\left/ {\vphantom {{\mathbb R} {\mathbb Z}}} \right.
\kern-\nulldelimiterspace} {\mathbb Z}  ,
\end{gather*}
one f\/inds that it is possible to associate with any 1-cycle $ C \in Z_1 \left( {M_3 } \right)$ a canonical DB class $\left[ {\eta _C } \right]\in {\widetilde H}_D^{ 1} \left( {M_3 ,{\mathbb Z}}
\right)$ such that
\begin{gather}
\label{24}
\exp \left\{ {2i\pi \int_C {\left[ \omega \right]} } \right\}=\exp \left\{
{2i\pi \int_{M_3 } {\left[ \omega \right]\ast \left[ {\eta _C } \right]} }
\right\}  ,
\end{gather}
for any $\left[ \omega \right]\in H_D^1 \left( {M_3 ,{\mathbb Z}} \right)$.   As an an alternative point of
view, consider a smoothing homotopy of $C$ within $H_D^1 \left(
{M_3 ,{\mathbb Z}} \right)$, that is,  a sequence of smooth DB classes
$\left[ {\eta _\varepsilon } \right]\in H_D^1 \left( {M,{\mathbb Z}} \right)$
such that (see \cite{HLZ03} for details)
\begin{gather}
\label{25}
\mathop {\lim }\limits_{\varepsilon \to 0}  \exp \left\{ {2i\pi \int_M
{\left[ A \right]\ast \left[ {\eta _\varepsilon } \right]} }
\right\}=\exp\left\{ {2i\pi \int_C {\left[ A \right]} } \right\}.
\end{gather}
This implies
\begin{gather}
\label{26}
\mathop {\lim }\limits_{\varepsilon \to 0}  \left[ {\eta _\varepsilon }
\right] =\left[ {\eta _C } \right]
\end{gather}
within the completion   ${\widetilde H}_D^{ 1} \left( {M_3 ,{\mathbb Z}}
\right)$ of $H_D^1 \left( {M_3 ,{\mathbb Z}} \right)$; this is why in \cite{HLZ03} $\left[ {\eta _C } \right]$  is said to belong to the boundary  of $H_D^1 \left( {M_3 ,{\mathbb Z}} \right)$. It should be noted that, by def\/inition, the limit (\ref{25}) and the corresponding limit (\ref{26}) are always well def\/ined. For this reason, in what  follows we shall concentrate directly to the distributional space ${\widetilde H}_D^{ 1} \left( {M_3 ,{\mathbb Z}} \right)$ and, in the presentation of the various arguments, the possibility of adopting a  limiting procedure of the type shown in equation~(\ref{25}) will  be simply understood.

Finally, let us point out that with the aforementioned geometrical interpretation of DB cohomology classes, the DB product of smooth classes canonically def\/ines a product within the space of Abelian Gerbes with connections. For instance, the DB product of two classes of $U(1)$-bundles with connections over $M$ turns out to be a class of $U(1)$-gerbe with connection over~$M$.

\subsection{Distributional forms and Seifert surfaces}\label{sec2.4}

How to construct the class $\left[ {\eta _C } \right] $, which enters equation (\ref{24}), is explained in detail for instance in   \cite{BGST05}. Here we outline the main steps of the construction and we consider, for illustrative purposes, the case $M_3 \sim S^3$. The integral  of a one-form $ \omega  $ along an oriented knot $C \subset S^3$ can be written as the integral on the whole $S^3$ of the external product $\omega \wedge J_C$, where the current~$J_C$ is a distributional 2-form with support on the knot $C$; that is, $\int_C \omega = \int_{S^3} \omega \wedge J_C$.  Since $J_C$ can be understood as the boundary of an oriented surface $\Sigma_C$ in $S^3$ (called a Siefert surface), one has $J_C= d \eta_C $ for some $1$-form $\eta_C$ with support on $\Sigma_C$.  One then f\/inds,
$\int_C \omega = \int_{S^3} \omega \wedge d \eta_C$, which  corresponds precisely to equation (\ref{24}) with $\left[ {\eta _C } \right]  \in  {\widetilde H}_D^{ 1} \left( {S^3 ,{\mathbb Z}}
\right)$ denoting the  Deligne cohomology class which is associated to $\eta_C $ and with $ [ \omega ] \in  H_D^{ 1} \left( {S^3 ,{\mathbb Z}} \right ) $ denoting the class which can be represented by $\omega $.

\begin{figure}[t]
\centerline{\includegraphics[width=70mm]{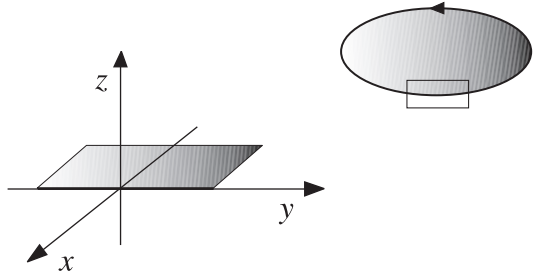}}
\caption{In a open domain with local coordinates $(x,y,z)$, a piece of a homologically
trivial loop $C$  can be identif\/ied with the $y$ axis, and the disc  that it
bounds (Seifert  surface) can be identif\/ied with a~portion of the half plane $(x<0$, $y,z=0)$.}
\label{fig2}
\end{figure}

 Let us consider, for instance, the unknot $C$ in $S^3$, shown in Fig.~\ref{fig2}, with a simple disc as Seifert surface. Inside the open domain depicted in Fig.~\ref{fig2}, the oriented knot is described~-- in local coordinates $(x,y,z)$~-- by a piece of the $y$-axis and the corresponding distributional forms~$J_C$ and~$\eta_C $ are given by
\begin{gather}
J_C = \delta (z)  \delta (x)  dz \wedge dx,   \qquad \eta_C = \delta (z)  \theta (-x) dz  .
\label{27}
\end{gather}

 For a generic 3-manifold $M_3$ and for each oriented knot $C \subset M_3$, the distributional 2-form~$J_C $ always exists, whereas a corresponding Seifert surface and the associated 1-form $\eta_C$ can in general be (globally) def\/ined only when the second cohomology group of $M_3$ is vanishing. Nevertheless, the class $[ \eta_C ]\in {\widetilde H}_D^{ 1} \left( {M,{\mathbb Z}} \right)$ is always well def\/ined for arbitrary 3-manifold $M_3$. In fact, when a Seifert surface associated with $C \subset M_3$ does not exist, the Chech--de~Rham cocycle  sequence representing $[ \eta_C ] \in {\widetilde H}_D^{ 1} \left( {M,{\mathbb Z}} \right)$ is locally of the form $( \eta_C^{(0,1)}, \Lambda_C^{(1,0)}, N_C^{(2,-1)})$ where,  inside suf\/f\/iciently small open domains, the expression of $ \eta_C^{(0,1)} $ is trivial or may coincide with the expression (\ref{27}) for $\eta_C$, and  $\Lambda_C^{(1,0)}$ and $N_C^{(2,-1)}$ are nontrivial components (when a~Seifert surface  exists,  the  components $\Lambda_C^{(1,0)}$ and $N_C^{(2,-1)}$ are trivial).

\section{Linking and self-linking}\label{sec3}

As we have already mentioned,  in the context of equation (\ref{24})  the pairing  $H_D^1 \left( {M_3 ,{\mathbb Z}} \right)\times {\widetilde H}_D^{ 1} \left( {M_3 ,{\mathbb Z}}
\right) \rightarrow {\widetilde H}_D^{ 3} \left( {M_3 ,{\mathbb Z}} \right)$  is well def\/ined. However, in what follows we shall also need to consider a  pairing induced by the DB product of the type
 ${\widetilde H}_D^1 \left( {M_3 ,{\mathbb Z}} \right)\times {\widetilde H}_D^{ 1} \left( {M_3 ,{\mathbb Z}} \right) \rightarrow {\widetilde H}_D^{ 3} \left( {M_3 ,{\mathbb Z}} \right)$ and this presents in general ambiguities that we need to f\/ix by means of some conventional procedure.

 \subsection{Linking number}\label{sec3.1}

 Let us consider f\/irst the case $M_3 \sim S^3$. Let $C_1$ and $C_2$ be two non-intersecting  oriented knots in~$S^3$ and let $\eta_1$ and $\eta_2$  the corresponding distributional 1-forms described in Section~\ref{sec2.4}, one has
\begin{gather}
 \int_{S^3}  \eta_1 \wedge d \eta_2  = \int_{S^3}  \eta_2 \wedge d \eta_1  = \ell k (C_1 , C_2) ,
 \label{28}
 \end{gather}
 where $\ell k (C_1 , C_2) $ denotes the linking number of~$C_1$ and~$C_2$, which is an integer valued ambient isotopy invariant. In fact,~$\eta_1 \wedge d \eta_2 $ represents an intersection form counting how many times~$C_2$ intersects the Seifert surface associated with $C_1$ (see also, for instance,~\cite{BT82, ROL}).
Let $ [\eta_1 ] $ and~$[ \eta_2 ]$ denote the DB classes which are associated with~$\eta_1$ and~$\eta_2$;   since the linking number is an integer, one f\/inds
 \begin{gather}
\exp \left \{  2i \pi \int_{S^3} [ \eta_1] *[  \eta_2 ] \right \} = \exp \left \{ 2i \pi \int_{S^3} [ \eta_2] *[  \eta_1 ] \right \} = \exp \left \{ 2i \pi \int_{S^3} \eta_1 \wedge d \eta_2 ] \right \} = 1.
 \label{29}
\end{gather}
Equations (\ref{28}) and (\ref{29}) show that the product $[\eta_1] * [\eta_2]$  is well def\/ined and just corresponds to the trivial class
\begin{gather}
[\eta_1] * [\eta_2] =  [0]  \in  {\widetilde H}_D^{ 3} \left( {S^3 ,{\mathbb Z}} \right) .
\label{30}
\end{gather}

In the next sections, we shall encounter the linking number in the DB cohomology context in the following form. Let $x$ be a real number, since $\eta_2 $ is globally def\/ined in $S^3$,  the 1-form $x \eta_2 $ is also globally def\/ined. Let us denote by $[ x \eta_2] $ the DB class which is represented by the form~$x \eta_2  $. One  has
\begin{gather}
 \exp \left \{ 2i \pi \int_{S^3} [ \eta_1] * [x  \eta_2 ]  \right \} = \exp \left \{ 2i \pi \int_{S^3}   \eta_1 \wedge d (x \eta_2) \right \} = \exp \left \{   2i \pi  x  \ell k  (C_1 , C_2)    \right \}  .
  \label{31}
 \end{gather}

\subsection{Framing}\label{sec3.2}

 Let  $\eta_C$ be the distributional 1-form which is associated with the oriented knot $C \subset S^3$; for a~single  knot,  the expression of  the self-linking number
  \begin{gather}
 \int_{S^3}  \eta_C \wedge d \eta_C
 \label{32}
\end{gather}
is not well def\/ined because the self-intersection form $ \eta_C \wedge d \eta_C $ has ambiguities. This means that, similarly to what happens with the product of distributions,
at the level of the class $[ \eta_C]  \in {\widetilde H}_D^1 \left( {S^3 ,{\mathbb Z}} \right)$, the product $[ \eta_C] \ast  [ \eta_C] $ is not well def\/ined a priori.

As shown in equations (\ref{25}) and (\ref{26}),  $[\eta_C] $ can be determined by means of the $\varepsilon \rightarrow 0 $ limit of $[ \eta_\varepsilon ] \in H_D^1 \left( {M_3 ,{\mathbb Z}} \right)$. One could then try to def\/ine the product $[ \eta_C] \ast  [ \eta_C] $ by means of the same limit
\begin{gather}
\lim\limits_{\varepsilon \to 0} \int_{S^3} [ \eta_\varepsilon ]  \ast [ \eta_\varepsilon ]  = \int_{S^3}  [ \eta_C] \ast  [ \eta_C]  .
\label{33}
\end{gather}
Unfortunately, the  limit  (\ref{33})  does not exist,  because the value that one obtains for the integral~(\ref{33}) in the $\varepsilon \rightarrow 0$ limit nontrivially depends on the way in which~$[ \eta_\varepsilon ] $ approaches $[ \eta_C ] $. This problem will be solved by the introduction of the framing procedure, which ultimately specif\/ies  how $[ \eta_\varepsilon ] $ approaches $[ \eta_C ] $. One should note that the ambiguities entering the integral (\ref{32}) and the limit (\ref{33}) also appear in the Gauss integral
\begin{gather}
\frac {1}{ 4 \pi} \oint_C dx^\mu \oint_C dy^\nu  \epsilon_{\mu \nu \rho}  \frac{( x- y)^\rho}{ | x-y |^3} ,
\label{34}
\end{gather}
which corresponds to the self-linking number.
A direct computation  \cite{SL} shows that the value of the integral (\ref{34}) is a real number which is not invariant under ambient isotopy transformations; in fact, it can be smoothly modif\/ied by means of smooth deformations of the knot $C$ in $S^3$. In order to remove all ambiguities and def\/ine the product $[\eta_C] \ast [ \eta_C ]$, we shall adopt the framing procedure  \cite{ROL, SLF}, which is also used for giving a topological meaning to the self-linking number.

\begin{definition}
\label{Definition3.2.1} A solid torus is a space homeomorphic to $S^1 \times D^2$, where $D^2$ is a two dimensional disc;  in the complex plane, $D^2$ can be represented  by the set $\{ z , {\rm ~with~}  |z|  \leq 1 \} $. Consider now an  oriented knot $C \subset S^3$; a tubular neighbourhood $V_C $ of $C $ is a solid torus embedded in $S^3$ whose core is $C $. A given homeomorphism $h : S^1\times D^2 \rightarrow  V_C$ is called a framing for $C $. The image of the standard longitude $h(S^1 \times 1)$ represents a knot $C_f \subset S^3$, also called the framing of $C $, which lies in a neighbourhood of $C $ and whose orientation is f\/ixed to agree with the orientation of $C $.  Up to isotopy transformations,  the homeonorphism $h$ is specif\/ied by
 $C_f$.
\end{definition}

 Clearly, the thickness of the  tubular neighbourhood $V_C $  of $C$ is chosen to be suf\/f\/iciently small so that, in the presence of several link components for instance, any knot  dif\/ferent from~$C$ belongs to the complement of $V_C \subset S^3 $.

For each framed knot $C $, with framing $C_f $,  the self-linking number of $C$ is def\/ined to be $\ell k (C  , C_f )$,
\begin{gather}
\int_{S^3} \eta_C \wedge d \eta_C  \equiv  \int_{S^3} \eta_C \wedge d \eta_{C_f }  =    \ell k (C , C_f ) .
\label{35}
\end{gather}

\begin{definition}
\label{Definition3.2.2} In agreement with equation (\ref{35}), one can consistently def\/ine the product \linebreak {$[\eta_C ] \ast [\eta_C]$} as
\begin{gather}
[ \eta_C ] \ast [\eta_C]   \equiv    [ \eta_C ] \ast [ \eta_{C_f} ]  .
\label{36}
\end{gather}
\end{definition}

 Def\/inition (\ref{36}) together with equations (\ref{35}) and (\ref{30}) imply that, for each framed knot $C$ (in $S^3$), the product $[ \eta_C] \ast [ \eta_C] $ is well def\/ined and corresponds to the trivial class
 \begin{gather*}
[\eta_C] * [\eta_C] =  [0]  \in  {\widetilde H}_D^{ 3} \left( {S^3 ,{\mathbb Z}} \right)  .
\end{gather*}

\begin{remark}
\label{Remark3.2.1} The product $[\eta_C ] \ast [\eta_C]$ also admits a def\/inition which dif\/fers from   equation (\ref{36}) but, as far as the computation of the Chern--Simons observables is concerned, is equivalent to equation (\ref{36}).  Instead of dealing with  a tubular neighbourhood $V_C $ with suf\/f\/iciently small but f\/inite thickness, one can def\/ine a limit in which  the transverse size of the neighbourhood $V_C $ vanishes. Let $\rho > 0 $ be the size of the diameter of the tubular neighbourhood $V_C (\rho )$ of the knot $C$; $\rho$ is def\/ined with respect to some (topology compatible) metric~$g$. The homeomorphism $h (\rho ) : S^1\times D^2 \rightarrow  V_C(\rho ) $ is assumed to depend smoothly on $\rho $. Then, the corresponding framing knot $C_f(\rho )$ also smoothly depends on $\rho $.
Consequently, the linking number $\ell k ( C , C_f (\rho )) $ does not depend on the value of $\rho $ and it will be denoted by $\ell k (C  , C_f)$. It should be noted that $\ell k (C  , C_f )$ does not depend on the choice of the metric $g$.
In the $\rho \rightarrow 0 $ limit, the solid torus $V_C (\rho ) $ shrinks to its core $C$ and the framing $C_f(\rho ) $ goes to $C$.  One can then def\/ine $\eta_C \wedge d \eta_C $ according to
\begin{gather}
\int_{S^3} \eta_C \wedge d \eta_C  \equiv   \lim \limits_{\rho \to 0}  \int_{S^3} \eta_C \wedge d \eta_{C_f (\rho )}  =  \lim \limits_{\rho \to 0}   \ell k (C , C_f (\rho ) ) =  \ell k (C , C_f ) .
\label {38}
\end{gather}
In agreement with equation (\ref{38}), one can put
\begin{gather}
[ \eta_C ] \ast [\eta_C]  \equiv  \lim \limits_{\rho \to 0}   [ \eta_C ] \ast [ \eta_{C_f (\rho )} ] .
\label {39}
\end{gather}
\end{remark}

\begin{remark}
\label{Remark3.2.2} The  def\/inition (\ref{36}) of  the DB product $[ \eta_C ] \ast [\eta_C] $ is consistent with equations (\ref{29})--(\ref{31}) and is topologically well def\/ined.  In fact, in the case of an oriented framed link~$L$ with $N$ components $\{ C_1 , C_2 ,\dots, C_N \}$ the corresponding canonical class $[\eta_L] \in  {\widetilde H}_D^{ 1} \left( {S^3 ,{\mathbb Z}} \right)  $ is equivalent to the sum of the classes which are associated with the single components, i.e.\  $[\eta_L] = \sum_j [\eta_j] $. Thus one f\/inds
\begin{gather}
 [\eta_L ] \ast [\eta_L]  =  \sum_j [\eta_j] \ast [\eta_j] + 2 \sum_{i < j} [\eta_i ] \ast [\eta_j] .
\label{40}
\end{gather}
The framing procedure which is used to def\/ine the DB product $ [\eta_L ] \ast [\eta_L]  $ guarantees that, if one integrates the 3-forms entering expression~(\ref{40}), the result does not depend on the particular choice of the Seifert surface which is used to (locally) def\/ine  the distributional forms associated with $L$.  This means that the framing procedure preserves both gauge invariance and ambient isotopy invariance.
\end{remark}

\begin{remark}
\label{Remark3.2.3} In order to def\/ine the extension of the DB product to distributional DB classes, one could try to start from equation (\ref{20bis}).   In this case,  the product of the DB representatives of two cycles (\ref{20bis}) would only contain local integral chains and the cup product $\ticup$ would just reduce to the intersection number of such integral chains (once these chains have been placed into transverse position, which is always possible because of  the freedom in the choice of the DB cocycles representing a given DB class). Accordingly, the extension of the product to the distributional case  would only produce integral chains and eventually integers in the integrals. Finally, by using smooth approximations of the cycles within (\ref{20bis}) and then performing the limits, as described above in equation (\ref{39}), one would obtain the same result. Note that,  in this last approach, the limit would be performed with the linking number  $\ell k (C, C_f ) $ f\/ixed. This is similar  to the def\/inition of the charge density of a charged point particle by taking the limit $r \rightarrow 0$ of a uniformly charged sphere of radius $r$ while keeping the total charge of the sphere f\/ixed, which leads to the well-known Dirac delta-distribution.

Knots or links can be framed in any oriented 3-manifold~$M_3$. In order to preserve the topological properties of the pairing ${\widetilde H}_D^1 \left( {S^3 ,{\mathbb Z}} \right)\times {\widetilde H}_D^{ 1} \left( {S^3 ,{\mathbb Z}} \right) \rightarrow {\widetilde H}_D^{ 3} \left( {S^3 ,{\mathbb Z}} \right)$ which is def\/ined by means of framing in $S^3$, we shall extend  the framing procedure to the case of a generic 3-manifold~$M_3$ by extending the validity of   properties  (\ref{30}) and  (\ref{36}).
\end{remark}

\begin{definition}
\label{Definition3.2.4}  If $[\eta_1] $ and $[\eta_2]$ are the classes in  ${\widetilde H}_D^{ 1} \left( {M_3 ,{\mathbb Z}}\right) $ which are canonically associated with the oriented nonintersecting knots $C_1$ and $C_2$ in $M_3$,  in agreement with equation (\ref{30}) we shall eliminate the (possible) ambiguities of the product  $[\eta_1] * [\eta_2] $ in such a way that
\begin{gather}
[\eta_1] * [\eta_2] =  [0]  \in  {\widetilde H}_D^{ 3} \left( {M_3 ,{\mathbb Z}} \right) .
\label{41}
\end{gather}
Consequently, for each oriented framed knot $C\subset M_3$ with framing $C_f$, we shall use the def\/inition
\begin{gather}
[ \eta_C ] \ast [\eta_C]   \equiv    [ \eta_C ] \ast [ \eta_{C_f} ] = [0]  \in
{\widetilde H}_D^{ 3} \left( {M_3 ,{\mathbb Z}}\right)   .
\label{42}
\end{gather}
\end{definition}

\begin{remark}
\label{Remark3.2.4} Def\/inition (\ref{42}) can also be understood by starting from equation (\ref{20bis}) and by using the same arguments that have been presented in the case $M_3 \sim S^3$. Let us point out that, unlike the $S^3$ case, for generic $M_3$ one f\/inds directly equation (\ref{42}) without the validity of some intermediate relations like equation (\ref{35}),  which may not be well def\/ined for $M_3 \not\sim S^3$.
\end{remark}

\section[Abelian Chern-Simons field theory]{Abelian Chern--Simons f\/ield theory}\label{sec4}

\subsection{Action functional}\label{sec4.1}

If one uses the Cech--de Rham double complex to describe DB classes, it can  easily be shown that the f\/irst component of a DB product of
a $U\left( 1 \right)$-connection $A$ with itself is given by
$A\wedge dA$ or, more precisely, it is given by the collection of all these products taken
in the open sets of a~good cover of $M_3$. This means that
the expression of the Chern--Simons lagrangian  of a $U\left( 1 \right)$-connection $A$ can be understood  as a DB class which coincides with the ``DB square'' of the class of $A$. Let  $\left[ A \right]$ denote the DB class associated to the $U\left( 1
\right)$-connection $A$,  the Chern--Simons functional  $S_{CS}$ is given by
\begin{gather*}
S_{CS} =  \int_{M_3 } {\left[ A \right]\ast \left[ A \right]} .
\end{gather*}
By def\/inition of the DB cohomology, the Chern--Simons action $S_{CS}$ is an element of ${\mathbb R} / {\mathbb Z}$ and then it is def\/ined modulo integers.
Consequently, in the functional measure  of the path-integral, the phase factor which is induced by the action  has to  be of the type
\begin{gather*}
\exp \left \{ 2 i \pi  k  S_{CS} \right \} =
\exp\left\{ {2i\pi k\int_{M_3 } {\left[ A \right]\ast \left[ A \right]} }
\right\} ,
\end{gather*}
where the coupling constant $k$ must be a nonvanishing integer
\begin{gather*}
k\in {\mathbb Z} , \qquad k \not=0 .
\end{gather*}
A modif\/ication of the orientation of $M_3$ is equivalent to the replacement $ k \rightarrow - k$.

\subsection{Observables}\label{sec4.2}

The observables that we shall consider are given by the expectation values of the Wilson line operators $W(L)$ associated with links $L$ in $M_3$. An oriented coloured and framed link $L \subset M_3$ with $N$ components is the union of non-intersecting knots $ \{ C_1 , C_2 , \dots , C_N \} $ in $M_3$, where each knot $C_j $ (with $j=1,2,\dots, N$) is oriented and framed, and its colour is represented by an integer charge $q_j \in  {\mathbb Z} $. For any given DB class $[A]$, the classical expression of $W(L)$  is given by
\begin{gather}
W(L) = \prod_{j=1}^N \exp \left \{ 2 i \pi q_j \int_{C_j} [ A ] \right \}
= \exp \left \{ 2 i \pi \sum_j q_j \int_{C_j} [ A ] \right \}  ,
\label{46}
\end{gather}
which actually corresponds to the pairing (\ref{11})
\begin{gather*}
W(L) = \exp\left\{ {2i\pi \int_L {\left[ A \right]} } \right\} \equiv \exp\left\{
{2i\pi \left\langle {\left[ A \right], L } \right\rangle_1 } \right\}  .
\end{gather*}
Once more, each factor
\begin{gather}
\label{48}
\exp\left\{ {2i\pi q_j \int_{C_j} {\left[ A \right]} } \right\}  ,
\end{gather}
which appears in expression (\ref{46}), is well def\/ined if and only if $q_j\in {\mathbb Z}$; that is why the charges associated with knots must take integer values.  A modif\/ication of the orientation of the knot~$C_j$ is equivalent to the replacement $q_j \rightarrow - q_j$. Obviously, any link component with colour  $q=0$  can be eliminated.

\begin{remark}
\label{Remark4.2.1} The classical expression   (\ref{46}) does not depend on the framing of the knots $\{ C_j \} $; however, only for framed links are the Wilson line operators  well def\/ined.   The point is that,
in the quantum Chern--Simons f\/ield theory, the f\/ield components correspond to distributional valued operators,  and the  Wilson line operators are formally  def\/ined  by  expression (\ref{46})  together with a set of specif\/ied rules which must be  used  to remove possible ambiguities in the computation of the expectation values.  In the operator formalism, these ambiguities are related to the product of f\/ield operators in the same point \cite{GMM, EG} and they are eliminated by means of a framing procedure.   In the  path-integral approach, we shall see that all the ambiguities are related to the def\/inition of the pairing ${\widetilde H}_D^1 \left( {M_3 ,{\mathbb Z}} \right)\times {\widetilde H}_D^{ 1} \left( {M_3 ,{\mathbb Z}} \right) \rightarrow {\widetilde H}_D^{ 3} \left( {M_3 ,{\mathbb Z}} \right)$; as it has been discussed in Section~\ref{sec3}, we shall use the framing of the link components to eliminate all ambiguities by means of the def\/inition (\ref{42}).
\end{remark}

\begin{remark}
\label{Remark4.2.2}  In equations (\ref{46}) and (\ref{48}), we have used the same symbol to denote knots and their homological representatives  because a canonical correspondence \cite{BT82} between them always exists.  At the classical level, for any integer $q$ one can identify the 1-cycle $q  C\in Z_1 (M) $ with the $q$-fold covering of the cycle $C$ or the $q$-times product of $C$ with itself. At the quantum level, this equivalence may not be valid when it is applied to the Wilson line operators because of ambiguities in the def\/inition of composite operators;  so, in order to avoid inaccuracies, we will always refer to Wilson line operators  def\/ined for oriented coloured and framed  knots or links.
\end{remark}

\begin{definition}
\label{Definition4.2.1} For each link component $C_j$, let $[ \eta_j ] \in {\widetilde H}_D^{ 1} \left( {M_3 ,{\mathbb Z}} \right) $ be the DB class such that
\begin{gather*}
\exp \left\{ {2i\pi q_j \int_{C_j} {\left[ A \right]} } \right\}=\exp \left\{
{2i\pi q_j \int_{M_3 } {\left[ A \right]\ast \left[ \eta _ j \right]} }
\right\} .
\end{gather*}
With the def\/inition
\begin{gather}
[ \eta_L ] = \sum_j q_j  [ \eta_j ]  ,
\label{50}
\end{gather}
one has
\begin{gather*}
\exp \left \{ {2 i \pi \int_{M_3} [ A ] *[ \eta_L ]} \right \} = \exp \left \{ {2 i \pi  \sum_j q_j \int_{M_3} [ A ] *[ \eta_j ]} \right  \} .
\end{gather*}
The expectation values of the Wilson line operators can be written in the form
\begin{gather}
\left\langle W(L)
\right\rangle _{k} \equiv   \frac{\int {D\left[ A \right]  \exp\left\{ {2i\pi
k\int_{M_3 } {\left[ A \right]\ast \left[ A \right]} } \right\}  W(L)  } }{\int {D\left[ A
\right]  \exp\left\{ {2i\pi k\int_{M_3 } {\left[ A \right]\ast \left[ A
\right]} } \right\}} } \nonumber \\
\phantom{\left\langle W(L)\right\rangle _{k}}{} = \frac{\int {D\left[ A \right] \exp\left\{ {2i\pi
k\int_{M_3 } {\left[ A \right]\ast \left[ A \right]} } \right\}  \exp\left\{
{2i\pi \int_{M_3 } {\left[ A \right]\ast \left[ {\eta _L } \right]} }
\right\}} }{\int {D\left[ A \right] \exp\left\{ {2i\pi k\int_{M_3 } {\left[ A
\right]\ast \left[ A \right]} } \right\}} } ,
\label{52}
\end{gather}
and our main purpose is to show how to compute them for arbitrary link $L$.
\end{definition}

\begin{remark}
\label{Remark4.2.3} In the DB cohomology approach, the functional integration (\ref{52})  locally corresponds to a sum over 1-form modulo forms with integer periods. So, the space of classical f\/ield conf\/igurations which are factorized out by gauge invariance  is in general larger than the standard  group of local gauge transformations.  It should be noted that this enlarged gauge symmetry perfectly f\/its the assumption that the expectation values (\ref{52}) form a  complete set of observables. In the DB cohomology interpretation of the functional integral  for  the quantum Chern--Simons f\/ield theory,  this enlargement of the ``symmetry group'' represents one of the main conceptual improvements with respect to the standard formulation of  gauge theories and, as we shall show, allows for a description of the functional space structure in terms of the homology groups of the manifold~$M_3$.
\end{remark}

\subsection {Properties of the functional measure}\label{sec4.3}

The sum over the DB classes $\int D [A] $ corresponds to a gauge-f\/ixed functional integral in ordinary quantum f\/ield theory, where one has to sum over the gauge orbits in the space of connections.   In the path-integral, smooth f\/ields conf\/igurations or f\/inite-action conf\/igurations have zero measure \cite{FH, COL}; so, the functional integral (\ref{52}) has to be understood as the functional integral in the appropriate extension or closure  ${\cal H}_D^{ 1} \left( {M_3 ,{\mathbb Z}} \right)$ of the space $ H_D^{ 1} \left( {M_3 ,{\mathbb Z}} \right)$, with  ${\widetilde H}_D^{ 1} \left( {M_3 ,{\mathbb Z}} \right) \subset  {\cal H}_D^{ 1} \left( {M_3 ,{\mathbb Z}} \right)$ and, more generaly, with
${\rm Hom}\left( {H_D^1 \left( {M,{\mathbb Z}}
\right),S^1} \right) \subset  {\cal H}_D^{ 1} \left( {M_3 ,{\mathbb Z}} \right)$.
In order to guarantee the consistency of the functional integral and its correspondence with ordinary gauge theories, we assume that the quantum measure  has the  following two properties.

\begin{enumerate}\itemsep=0pt
\item[({\bf M1})] {\it The space $ {\cal H}_D^{ 1} \left( {M_3 ,{\mathbb Z}} \right)$  inherits its structure from $ H_D^{ 1} \left( {M_3 ,{\mathbb Z}} \right)$ in agreement with se\-quen\-ce~\eqref{8}}.
\end{enumerate}

Equation (\ref{8}) then implies that the sum over DB classes is locally equivalent to a sum  over  $  {\Omega ^1\left( {M_3 } \right)} /  {\Omega _{\mathbb Z}^1 \left( {M_3 } \right)} $, i.e.\ a sum over
1-forms modulo generalized gauge transformations.

\begin{enumerate}\itemsep=0pt
\item[({\bf M2})]  {\it The functional measure is translational invariant.}
\end{enumerate}

This implies in particular that, for any $[\omega ] \in {\widetilde H}_D^{ 1} \left( {M_3 ,{\mathbb Z}} \right)$,  the quadratic  measure
\begin{gather}
\label{53}
d\mu _k \left( {\left[ A \right]} \right)\equiv D\left[ A \right] \exp\left\{
{2i\pi k\int_{M_3 } {\left[ A \right]\ast \left[ A \right]} } \right\}
\end{gather}
satisf\/ies the equation
\begin{gather}
\label{54}
d\mu _k \left( {\left[ A \right] + \left[ \omega \right]} \right) = d\mu _k \left(
{\left[ A \right]} \right)   \exp\left\{ {4i\pi k\int_{M_3 } {\left[ A
\right]\ast \left[ \omega \right]} + 2i\pi k\int_{M_3 } {\left[ \omega
\right]\ast \left[ \omega \right]} } \right\},
\end{gather}
which looks like a Cameron--Martin formula (see for instance \cite{ET84} and references therein).

Equation (\ref{54}) will be used extensively in our computations. The importance of generalized Wiener measures in the functional integral~-- which necessarily imply the validity of the Cameron--Martin property~-- and of the singular connections was also stressed in the articles   \cite{AL93} and \cite{Ba93} in which, however, the space of the functional integral is supposed to coincide with the space  of the classes of smooth connections on a f\/ixed $U(1)$-bundle over $M_3$.

In the computation of the observables (\ref{52}), we shall not use perturbation theory; only properties ({\bf M1}) and ({\bf M2}) of the functional measure will be utilized. We shall now derive the main properties of the observables of the Abelian Chern--Simons theory which are valid for any 3-manifold $M_3$.

\subsection{Colour periodicity}\label{sec4.4}

The colour of each oriented knot or link component  $C \subset M_3$ is specif\/ied  by the value of its  associated charge $q \in \mathbb Z$. For f\/ixed nonvanishing value of the coupling constant $k$, the expectation values (\ref{52}) are invariant under the substitution $q \rightarrow q + 2k $,  where $q$ is the charge of a generic link component. Consequently, one has

\begin{proposition}
\label{Proposition4.4.1} For fixed integer $k$,  the colour space is given by ${\mathbb Z}_{2k}$ which coincides with the space of residue classes of integers {\rm mod} $2k$.
\end{proposition}

\begin{proof} Let $\{ q_j \} $  be the charges which are associated with the components $\{ C_j\} $  ($j=1,2,\dots,N$) of the  link $L$. With the notation  (\ref{53}), the expectation value $\left\langle W(L)
\right\rangle _{k}$ shown in equation (\ref{52}) can be written as
\begin{gather}
\left\langle W(L)
\right\rangle _{k}  =   \frac{\int d\mu _k  (
[ A ] )   \exp \left \{ {2 i \pi  \sum_j q_j \int_{M_3} [ A ] *[ \eta_j ]} \right  \}   } {\int d\mu _k  ( [ A ] ) }  .
\label{55}
\end{gather}
Property ({\bf M2}) implies that, with the substitution $[A] \rightarrow [A] + [\eta_1 ] $,  the numerator of expres\-sion~(\ref{55}) becomes
 \begin{gather*}
  \int d\mu _k  (
[ A ] )   \exp \left \{ {2 i \pi  \sum_j q_j \int_{M_3} [ A ] *[ \eta_j ]} \right \} =
  \int d\mu _k  (
[ A ] )  \exp \left \{ {2 i \pi  \sum_j q^\prime_j \int_{M_3} [ A ] *[ \eta_j ]} \right \}
\nonumber \\
\qquad{}  \times
\exp \left \{  2i\pi k \int_{M_3} [\eta_1 ] \ast [\eta_1] \right \}
  \exp \left \{  2i \pi \sum_j q_j \int_{M_3} [\eta_1 ] \ast [\eta_j] \right  \}    ,
\end{gather*}
where $q^\prime_j = q_j + 2 k  \delta_{j1}$. In agreement with equation (\ref{41}), for $j \not= 1$ one has $[ \eta_1] \ast [\eta_j] \simeq  [0] \in  {\widetilde H}_D^{ 3} \left( {M_3 ,{\mathbb Z}} \right)$, and then
\begin{gather*}
\exp \left \{  2i \pi q_j \int_{M_3} [\eta_1 ] \ast [\eta_j] \right  \} = 1  .
\end{gather*}
Similarly, in agreement with equation (\ref{42}), by means of the framing procedure one obtains  $[ \eta_1] \ast [\eta_1] \simeq [0] \in  {\widetilde H}_D^{ 3} \left( {M_3 ,{\mathbb Z}} \right)$, and then
\begin{gather*}
\exp \left \{  2i \pi (q_1 + k)  \int_{M_3} [\eta_1 ] \ast [\eta_1] \right  \} = 1  .
\end{gather*}
Consequently, the numerator of expression (\ref{55}) can be written as
  \begin{gather*}
  \int d\mu _k  (
[ A ] )   \exp \left \{ {2 i \pi  \sum_j q_j \int_{M_3} [ A ] *[ \eta_j ]} \right \} \nonumber\\
\qquad{}=  \int d\mu _k  (
[ A ] )  \exp \left \{ {2 i \pi  \sum_j q^\prime_j \int_{M_3} [ A ] *[ \eta_j ]} \right \}
 ,
\end{gather*}
which proves that, for f\/ixed $k$,  the expectation values (\ref{52}) are invariant under the substitution $q_1 \rightarrow q_1 + 2k $,  where $q_1$ is the charge of the link component $C_1$.
\end{proof}

\subsection{Ambient isotopy invariance}\label{sec4.5}

Two oriented framed coloured links $L$ and $L^\prime$ in $M_3$ are ambient isotopic if $L$ can be smoothly connected with $L^\prime$ in $M_3$.

\begin{proposition}
\label{Proposition4.5.1} The Chern--Simons expectation values \eqref{52} are invariants of  ambient isotopy for framed links.
\end{proposition}

\begin{proof} Suppose that two  oriented  coloured framed links  $L $ and $L^\prime $ are ambient isotopic in $M_3$. The link $L$ has components $\{ C_1 , C_2, \dots, C_N\} $ with colours $ \{ q_1, q_2,\dots, q_N \}$; whereas the link $L^\prime $  has   components $\{ C^{ \prime}_1 , C_2 ,\dots, C_N \}$ with colours $\{ q_1 , q_2 ,\dots, q_N\}$, so that
\begin{gather}
[\eta_L ] = q_1  [ \eta_1 ] + \sum_{j\not= 1} q_j  [ \eta_j ] , \qquad
[\eta_{L^\prime }  ] = q_1  [ \eta^{ \prime}_1  ] + \sum_{j\not= 1} q_j  [ \eta_j ]  ,
\label{60}
\end{gather}
where the class $[ \eta_1 ] $ refers to the knot $C_1 \subset M_3$ and $ [ \eta_1^{ \prime }  ] $ is associated to the knot $C_1^{ \prime } \subset M_3$.

Let  $\tau : [0,1] \rightarrow C_1 (\tau ) \subset M_3$ be the isotopy which connects $C_1$ with $C_1^{ \prime} $ in $M_3$, with  $C_1 (0) = C_1 $ and $C_1(1)= C_1^{ \prime} $. We shall denote by $\Sigma \subset M_3 $ the 2-dimensional surface which has support on $\{ C_1 (\tau ) \subset M_3  ;  0 \leq \tau \leq 1 \}$; because of the freedom in the choice of $\tau $ within the same ambient isotopy class,  it is assumed that $\Sigma $ is well def\/ined and presents no singularities.   $\Sigma $ belongs to the complement  of the link components  $ \{ C_2 , C_3,\dots, C_N \}$ in $M_3$ and one can introduce an orientation on $\Sigma $ in such a way that  its oriented boundary is given by  $\partial  \Sigma = C_1^{ \prime} \cup C^{-1}_1 $, where $C^{-1}_1$ denotes the  knot  $C_1$ with reversed orientation.

The distributional 1-form $\eta_{\Sigma } $, which is associated with $\Sigma $, is globally def\/ined in $M_3$ and satisf\/ies
\begin{gather}
d  \eta_{ \Sigma } = d \eta_1^{\prime } - d \eta_1  .
\label{61}
\end{gather}
where, with a small abuse of notation, $d \eta_1$ and $d \eta_1^{ \prime }$ denote the integration currents of $C_1$ and~$C_1^{\prime}$ respectively. For $j \not= 1 $ one f\/inds
\begin{gather}
\int_{M_3} \eta_{\Sigma} \wedge d  \eta_j  = 0  ,
\label{62}
\end{gather}
because the link components $\{ C_2, C_3,\dots, C_N \}$ do not intersect the surface~$\Sigma $. Moreover, according to the framing procedure, the orientation of $\Sigma $ implies
\begin{gather}
\int_{M_3} \eta_{\Sigma} \wedge (d \eta_1^{ \prime} + d \eta_1 ) =
 \int_{{C_{1f}^{\prime}}}  \eta_{\Sigma}    +  \int_{{C_{1f}}}  \eta_{\Sigma} =0  ,
 \label{63}
\end{gather}
where ${C_{1f}^{\prime}} $ denotes the framing of $C_1^{ \prime}$ and ${C_{1f}}$ represents the framing of $C_1$.
Since $\eta_{ \Sigma }$ is globally def\/ined in $M_3$, the 1-form  $ x   \eta_{ \Sigma }  $  (with $x = (q_1 / 2 k) \in {\mathbb R}$)  is also globally def\/ined. Let $[ x \eta_{ \Sigma } ] \in {\widetilde H}_D^{ 1} \left( {M_3 ,{\mathbb Z}} \right)$ be the DB class which can be represented by the 1-form $x  \eta_{ \Sigma } $;  by construction, one has
\begin{gather}
\exp \left \{ 4 i \pi k  \int_{M_3} [ A ] *[ (q_1 / 2 k) \eta_{ \Sigma } ] \ \right \}  \nonumber \\ \qquad{} =
\exp \left \{  2 i \pi q_1  \int_{M_3} [ A ] *[ \eta_1^\prime  ]  \right \}
\exp \left \{ - 2 i \pi q_1    \int_{M_3} [ A ] *[ \eta_1 ]  \right \}  .
\label{64}
\end{gather}
The expectation value $\left\langle W(L) \right\rangle _{k} $ is given by
\begin{gather}
\left\langle W(L)
\right\rangle _{k}  =   \frac{\int d\mu _k  (
[ A ] )   \exp \left \{ {2 i \pi  \int_{M_3} [ A ] *[ \eta_L ]} \right  \}   } {\int d\mu _k  (
[ A ] ) } .
\label{65}
\end{gather}
Equation (\ref{64}) and property ({\bf M2})  imply that, with the substitution $[A] \rightarrow [A] + [ x \eta_{ \Sigma } ] $, the numerator of expression (\ref{65}) can be written as
\begin{gather*}
\int d\mu _k  ([ A ] )   \exp \left \{ 2 i \pi  \int_{M_3} [ A ] *[ \eta_{L^\prime } ] \right  \}   \nonumber \\
\qquad{} \times
 \exp \left \{ 2 i \pi k  \int_{M_3}[ x \eta_{ \Sigma } ]  \ast [ x \eta_{ \Sigma } ]  \right  \}
 \exp \left \{ 2 i \pi   \int_{M_3}[ x \eta_{ \Sigma } ]  \ast [  \eta_L ]  \right  \}  .
\end{gather*}
By using the relations
\begin{gather*}
 \exp \left \{ 2 i \pi k  \int_{M_3}[ x \eta_{ \Sigma } ]  \ast [ x \eta_{ \Sigma } ]  \right  \} =
  \exp \left \{  ( i  \pi q_1^2 /  2k)  \int_{M_3} \eta_{ \Sigma } \wedge ( d \eta_1^{ \prime } - d \eta_1 )    \right  \}  ,
\\
 \exp \left \{ 2 i \pi   \int_{M_3}[ x \eta_{ \Sigma } ]  \ast [  \eta_L ]  \right  \}  =  \exp \left \{  ( i  \pi q_1^2 /  k)  \int_{M_3} \eta_{ \Sigma } \wedge  d \eta_1     \right  \}    \nonumber \\
\qquad{}
\times   \exp \left \{ ( i \pi q_1 / k ) \sum_{j\not= 1} q_j  \int_{M_3}  \eta_{ \Sigma } \wedge d   \eta_j  \right  \}  ,
\end{gather*}
and equations (\ref{61})--(\ref{63}), one f\/inds that the numerator of expression   (\ref{65}) assumes the form
\begin{gather*}
\int d\mu _k  ([ A ] )   \exp \left \{ 2 i \pi  \int_{M_3} [ A ] *[ \eta_{L^\prime } ] \right  \}  .
\end{gather*}
Consequently,  the expectation values of the Wilson line operators  associated with the links~$L$ and~$L^\prime $, entering equation (\ref{60}), are equal. The same argument, applied to all the link components,  implies that, for any two ambient isotopic links $L$ and $L^\prime$, one has
\begin{gather*}
\left\langle W(L)
\right\rangle _{k}  =  \left\langle W(L^\prime )
\right\rangle _{k}  .
\end{gather*}
This concludes the proof.
\end{proof}

\subsection{Satellite relations}\label{sec4.6}

For the  oriented framed knot  $C\subset M_3$, let the homeomorphism $h : S^1\times D^2 \rightarrow  V_C$ be the framing of $C$, where $V_C$ is a
a tubular neighbourhood of $C $.  Let us represent the disc $D^2$ by the set  $\{ z , {\rm ~with~}  |z|  \leq 1 \} $ of the complex plane.  The framing $C_f$ of $C$ is given by $ h (S^1 \times 1)$, whereas one can always imagine that the knot $C$ just corresponds to $h ( S^1 \times 0) $. Let $P$ be a link in the solid torus  $S^1\times D^2$;  if one replaces the knot $C\subset M_3$ by $h (P) \subset M_3$ one obtains the satellite of~$C$ which is def\/ined by the pattern link $P$.

\begin{definition}
\label{Definition4.6.1} Let  $B  \subset S^1\times D^2$ be the oriented link with two components $ \{ B_1 , B_2 \} $ given by $B_1 = (S^1 \times 0) \subset S^1 \times D^2$ and $B_2 = (S^1 \times 1/2) \subset S^1 \times D^2 $. For any oriented framed knot $C \subset M_3$, let us denote by $C^{(2)} \in M_3$ the satellite of $C$ with is obtained by means of  the pattern link $B$.  The two oriented components $\{ K_1 , K_2 \}$ of $C^{(2)}$ are given by $K_1 = h (B_1 ) $ and $K_2= h (B_2)$. Let us introduce a framing for the components of the link $C^{(2)}$; the knot  $K_1 $ has framing $ K_{1f} = h(S^1 \times 1/4)$ and the knot $K_2 $ has framing $ K_{2f} = h(S^1 \times 3/4)$.

By construction, the satellite $C^{(2)}$ of $C$ is an oriented framed link.
\end{definition}

\begin{proposition}
\label{Proposition4.6.1} Let $L$ and $\widetilde L$ be two  oriented coloured framed links in $ M_3$ in which  $ {\widetilde L } $ is obtained from $L = \{ C_1, \dots, C_N \} $ by substituting the component $C_1$, which has colour $q_1 \in \mathbb Z$,  with its satellite $C_1^{(2)}$ whose components $K_1$ and $K_2$ have colours ${\widetilde q}_1 = q_1  \pm 1$ and  ${\widetilde q}_2 =  \mp 1$ respectively. Then,  the corresponding Chern--Simons expectation values satisfy
\begin{gather}
\langle W(L) \rangle_k = \langle W({\widetilde L}) \rangle_k   .
\label{71}
\end{gather}
\end{proposition}

\begin{proof} Because of the ambient isotopy invariance of
$ \langle W({\widetilde L}) \rangle_k $, one can consider the limit in which the  component $K_1$  approaches to $K_2$ and  coincides with $K_2$. In this limit,   for  each f\/ield conf\/iguration (i.e.\ for each DB class) the associated holonomies $W(C_1) $ and $W(C^{(2)}_1 )$ coincides. This means that,  at the classical level,
equality (\ref{71}) is satisf\/ied. Thus, we only need to consider  possible ambiguities in the expectation value of the  composite Wilson line operator $W(C^{(2)}_1 )= W(K_1) W(K_2) $ in the $K_1 \rightarrow K_2 $ limit. In agreement with what we shall show in the following sections, we now assume that all the ambiguities which refer to composite Wilson line operators are eliminated by means of the framing procedure which is used to def\/ine the product $[\eta_{\widetilde L}] \ast [\eta_{\widetilde L} ]$.
According to the def\/inition (\ref{50}), one has
\begin{gather*}
  [\eta_L ] = q_1 [\eta_1] + \sum_{j=2}^N q_j [\eta_j] = q_1 [\eta_1 ] + [ {\overline \eta }_L ]  ,
\\
  [ \eta_{{\widetilde L}} ]  =  {\widetilde q}_1 [\eta_{K_1}] + {\widetilde q}_2 [\eta_{K_2}] +  \sum_{j=2}^N q_j [\eta_j] =  {\widetilde q}_1 [\eta_{K_1}] + {\widetilde q}_2 [\eta_{K_2}] + [ {\overline \eta }_L ]  ,
\end{gather*}
 and then
\begin{gather*}
[\eta_L ] \ast [\eta_L ] = q_1^2 [\eta_{C_1}] \ast [\eta_{C_{1}} ]  + 2 q_1 [\eta_{C_1} ] \ast [ {\overline \eta }_L ] + [ {\overline \eta }_L ] \ast [ {\overline \eta }_L ]  , 
\\
[\eta_{\widetilde L} ] \ast [\eta_{\widetilde L} ]   = \left ( {\widetilde q}_1 [ \eta_{K_1}] + {\widetilde q}_2  [\eta_{K_2}] \right ) \ast
\left ( {\widetilde q}_1 [ \eta_{K_1}] + {\widetilde q}_2  [\eta_{K_2}] \right ) \nonumber \\
\phantom{[\eta_{\widetilde L} ] \ast [\eta_{\widetilde L} ]   =}{} + 2 \left ( {\widetilde q}_1 [ \eta_{K_1}] + {\widetilde q}_2  [\eta_{K_2}] \right )
\ast [ {\overline \eta }_L ] + [ {\overline \eta }_L ] \ast [ {\overline \eta }_L ]  .  
\end{gather*}
As far as the computation of the Chern--Simons observables is concerned,
ambient isotopy invariance and equality $q_1 = {\widetilde q}_1 + {\widetilde q}_2 $ imply
\begin{gather*}
2 q_1 [\eta_{C_1} ] \ast [ {\overline \eta }_L ] = 2 \left ( {\widetilde q}_1 [ \eta_{K_1}] + {\widetilde q}_2  [\eta_{K_2}] \right )
\ast [ {\overline \eta }_L ]  ,
\end{gather*}
moreover, by construction of the satellite $C_1^{(2)}$ and the def\/inition (\ref{42}), one also f\/inds
\begin{gather*}
q_1^2 [\eta_{C_1}] \ast [\eta_{C_{1}} ]  = \left ( {\widetilde q}_1 [ \eta_{K_1}] + {\widetilde q}_2  [\eta_{K_2}] \right ) \ast  \left ( {\widetilde q}_1 [ \eta_{K_1}] + {\widetilde q}_2  [\eta_{K_2}] \right )  .
\end{gather*}
Therefore,  as far as the computation of the Chern--Simons observables is concerned, one can replace $[\eta_L ] \ast [\eta_L ] $ by $[\eta_{\widetilde L} ] \ast [\eta_{\widetilde L} ] $, and then $\langle W(L) \rangle_k = \langle W({\widetilde L}) \rangle_k$.
\end{proof}

\begin{definition}
\label{Definition4.6.2}  In agreement with Proposition~\ref{Proposition4.6.1}, for any oriented coloured framed link  $L\subset M_3$,  one can replace recursively all the link components which have colour given by $q \not= \pm 1$ by their satellites constructed with the pattern link $B$,  in such a way that the resulting link ${\overline L}\subset M_3 $ has the following property:  each oriented framed component of ${\overline L}$ has colour  which is specif\/ied by a charge $q = + 1$ or $q=-1$.  Remember that, for each  link component $C$,  the sign of the associated charge $q$ is def\/ined with respect to the orientation of $C$.  So, with a suitable choice of the orientation of the link components, all the link components of  ${\overline L}$  have  charges~$+1$.
For each link $L\subset M_3$, the corresponding link  ${\overline L}\subset M_3 $ will be called the  {\it simplicial satellite} of $L$ and, as a consequence of Proposition~\ref{Proposition4.6.1}, one has
\begin{gather}
\langle W(L) \rangle_k = \langle W({\overline L}) \rangle_k   .
\label{78}
\end{gather}
\end{definition}

\section[Abelian Chern-Simons theory on $S^3$]{Abelian Chern--Simons theory on $\boldsymbol{S^3}$}\label{sec5}

When  $M_3 = S^3$, the DB cohomology group satisf\/ies $H_D^1 \left( {S^3,{\mathbb Z}} \right)\simeq {\Omega^1\left( {S^3} \right)} \mathord{\left/ {\vphantom {{\Omega ^1\left( {S^3} \right)} {\Omega _{\mathbb Z}^1 \left( {S^3} \right)}}} \right.
\kern-\nulldelimiterspace} {\Omega _{\mathbb Z}^1 \left( {S^3} \right)}$ and one  has ${\Omega^1\left( {S^3} \right)} \mathord{\left/ {\vphantom {{\Omega ^1\left( {S^3} \right)} {\Omega _{\mathbb Z}^1 \left( {S^3} \right)}}} \right.
\kern-\nulldelimiterspace} {\Omega _{\mathbb Z}^1 \left( {S^3} \right)}={\Omega^1\left( {S^3} \right)} \mathord{\left/ {\vphantom {{\Omega ^1\left( {S^3} \right)} {d\Omega ^0\left( {S^3} \right)}}} \right. \kern-\nulldelimiterspace} {d\Omega ^0\left( {S^3} \right)}$.
Since in general the path-integral of the Chern--Simons theory on $M_3$ locally corresponds to a sum over the space of 1-forms modulo forms with integer periods,  it is convenient to introduce a new notation; with respect to the origin of  $ {\Omega^1\left( {S^3} \right)} \mathord{\left/ {\vphantom {{\Omega ^1\left( {S^3} \right)} {\Omega _{\mathbb Z}^1 \left( {S^3} \right)}}} \right.
\kern-\nulldelimiterspace} {\Omega _{\mathbb Z}^1 \left( {S^3} \right)}$
that one can choose to correspond to the vanishing connection, an element of $ {\Omega^1\left( {S^3} \right)} \mathord{\left/ {\vphantom {{\Omega ^1\left( {S^3} \right)} {\Omega _{\mathbb Z}^1 \left( {S^3} \right)}}} \right.
\kern-\nulldelimiterspace} {\Omega _{\mathbb Z}^1 \left( {S^3} \right)}$ will be denoted by $[\alpha ]$.  So that, in agreement with  property  ({\bf M1}), for any oriented coloured and framed link $L \subset S^3$ the expectation value  (\ref{52})  can be written as
\begin{gather}
\left\langle W(L)
\right\rangle _{k} =  \frac{\int {D\left[ \alpha \right]  \exp\left\{ {2i\pi
k\int_{S^3 } {\left[ \alpha \right]\ast \left[ \alpha \right]} } \right\}  \exp\left\{
{2i\pi \int_{S^3 } {\left[ \alpha \right]\ast \left[ {\eta _L } \right]} }
\right\}} } {\int {D\left[ \alpha \right]  \exp\left\{ {2i\pi k\int_{S^3 } {\left[ \alpha \right]\ast \left[ \alpha
\right]} } \right\}} } \nonumber \\
\phantom{\left\langle W(L) \right\rangle _{k}}{} = \frac {\int d \mu_k ( [\alpha ])   \exp\left\{
2i\pi \int_{S^3 } \left[ \alpha \right] \ast \left[ \eta _L  \right] \right\} }{ \int d \mu_k ( [\alpha ]) },
\label{79}
\end{gather}
where  $[\alpha ] \in {\Omega^1\left( {S^3} \right)} / {\Omega _{\mathbb Z}^1 \left( {S^3} \right)}$  and
$[\eta_L] \in {\widetilde H}_D^{ 1} \left( {M_3 ,{\mathbb Z}} \right)$ denotes the class which is canonically associated with $L $.  The integral (\ref{79}) actually extends to ${\cal H}_D^{ 1} \left( {S^3 ,{\mathbb Z}} \right)$ which has to be understood as a suitable extension of ${\Omega^1\left( {S^3} \right)} / {\Omega _{\mathbb Z}^1 \left( {S^3} \right)}$. We shall now compute the observable $ \langle W(L) \rangle_k$ for arbitrary link $L$.

\begin{theorem}
\label{Theorem1}  Let the oriented coloured and framed link components  $\{ C_j \} $  of the link $L$,   with $j=1,2,\dots,N$, have charges  $\{ q_j\} $ and framings $ \{ C_{jf} \}$.  Then
\begin{gather}
\langle W(L) \rangle _{k} =  \exp \left \{ -( 2 i \pi / 4k)  \sum_{ij} q_i  {\mathbb L}_{ij} q_j  \right  \}   ,
\label{89a}
\end{gather}
 where the linking matrix $ {\mathbb L}_{ij}$ is defined by
\begin{gather*}
 {\mathbb L}_{ij} = \int_{S^3} \eta_i \wedge d  \eta_j = {\ell} k ( C_i , C_j ) , \qquad \mbox{for}  \quad i \not= j,
\\
 {\mathbb L}_{jj} = \int_{S^3} \eta_j \wedge d  \eta_j = {\ell} k ( C_j , C_{jf} )  .
\end{gather*}
\end{theorem}

\begin{proof} Since $H^2\left( {S^3,{\mathbb Z}} \right)=0$, Poincar\'{e} duality implies that any 1-cycle on $S^3$ is homologically trivial. Equivalently, for each knot $C_j$ one can f\/ind  an oriented Seifert surface $\Sigma_j \subset S^3$  such that $\partial  \Sigma_j = C_j $ (in fact, there is an inf\/inite number of topologically inequivalent Seifert surfaces) and one can then def\/ine a distributional 1-form $\eta_j$ (with support on $\Sigma_j$)  which is globally def\/ined in~$S^3$.  The distributional 1-form $\eta_L$  associated with the link $L$,
\begin{gather*}
\eta_L = \sum_j q_j  \eta_j  ,
\end{gather*}
is globally def\/ined in $S^3$ and, in the Chech--de Rham description of DB cocycles, the class $[\eta_L]$ can be represented by the sequence $( \eta_L , 0 , 0)$. The distributional  1-form
\begin{gather*}
\eta_L / 2k = \sum_j (q_j / 2k) \eta_j
\end{gather*}
is also globally def\/ined in $S^3$ and we shall denote by $[\eta_L / 2k ]\in {\widetilde H}_D^{ 1} \left( {M_3 ,{\mathbb Z}} \right)$ the DB class which, in the Chech--de Rham description of DB cocycles, is represented by the sequence $( \eta_L / 2k , 0 , 0 )$.
It should be noted that the class  $[\eta_L / 2k ]$ does not depend on the particular choice  of the 1-form~$\eta_L$ which represents $[\eta_L]$. (In turn, this implies that $[\eta_L / 2k ]$ does not depend on the particular choice of the Seifert surfaces.)  In fact, any representative 1-form of $[\eta_L]$ can be written as  $\eta_L + d \chi $ for some $\chi \in \Omega^0 (S^3)$; therefore, for the corresponding  class $[( \eta_L + d \chi ) / 2k ] $ one f\/inds
\begin{gather*}
[( \eta_L + d \chi ) / 2k ]= [ \eta_L / 2k  + d \chi /2k ] = [\eta_L /2k ] + [ d ( \chi / 2k)] = [\eta_L / 2k]  .
\end{gather*}
By construction, the class $ [ \eta _L / 2k ]$ satisf\/ies the relation
\begin{gather*}
2k [ \eta_L / 2k ] = [ \eta_L ]  ,
\end{gather*}
therefore
\begin{gather}
\exp \left\{ 4i\pi k \int_{S^3 } [ \alpha ] \ast  [ \eta_L / 2k ] \right \} =
\exp\left\{ 2i\pi \int_{S^3 } [ \alpha] \ast [ \eta_L ] \right \}  .
\label{84}
\end{gather}
In agreement with property ({\bf M2}), by means of the substitution $ [\alpha ] \rightarrow [\alpha ] -  [ \eta_L / 2k ] $ the numerator of expression (\ref{79}) assumes the form
\begin{gather}
\int d\mu _k  ([ \alpha ] )   \exp \left \{ - 4 i \pi k  \int_{S^3} [ \alpha ] *[ \eta_L / 2k ] \right  \}
 \exp \left \{ 2 i \pi k  \int_{S^3} [ \eta_L / 2k  ] *[ \eta_L / 2k ] \right  \}    \nonumber \\
\qquad{}{} \times
 \exp \left \{ 2 i \pi k  \int_{S^3}[ \alpha ]  \ast [  \eta_L ]  \right  \}
 \exp \left \{ - 2 i \pi   \int_{S^3}[  \eta_L / 2k  ]  \ast [  \eta_L ]  \right  \}  .
\label{85}
\end{gather}
With the help of equation (\ref{84}),  expression (\ref{85}) becomes
\begin{gather*}
\exp \left \{ -( 2 i \pi / 4k)   \int_{S^3} \eta_L  \wedge d  \eta_L \right  \}  \int d\mu _k  ([ \alpha ] )  ,
\end{gather*}
and then
\begin{gather*}
\langle W(L) \rangle _{k} = \exp \left \{ -( 2 i \pi / 4k)   \int_{S^3} \eta_L  \wedge d  \eta_L \right  \}  \frac{ \int d\mu _k  ([ \alpha ] )}{ \int d\mu _k  ([ \alpha ] )}  .
\end{gather*}
Assuming that, for  the manifold $S^3$, one has
\begin{gather*}
\int d\mu _k  ([ \alpha ] ) \not= 0  ,
\end{gather*}
one f\/inally obtains
\begin{gather}
\langle W(L) \rangle _{k} = \exp \left \{ -( 2 i \pi / 4k)   \int_{S^3} \eta_L  \wedge d  \eta_L \right  \}\nonumber\\
\phantom{\langle W(L) \rangle _{k}}{} = \exp \left \{ -( 2 i \pi / 4k)  \sum_{ij} q_i q_j  \int_{S^3} \eta_i  \wedge d  \eta_j \right  \}   ,
\label{89}
\end{gather}
which coincides with expression(\ref{89a}); and this concludes the proof.
\end{proof}

\begin{remark}
 \label{Remark5.1}  Expression (\ref{89a})  describes an invariant of ambient isotopy  (Proposition~\ref{Proposition4.5.1}) for oriented coloured framed links.
Since the matrix  elements  $ {\mathbb L}_{ij}$ are integers, in agreement with Proposition~\ref{Proposition4.4.1} the observable  (\ref{89a}) is invariant under the substitution $q_i \rightarrow q_i + 2k $ (for f\/ixed~$i$). Moreover, one can verify that Proposition~\ref{Proposition4.6.1} is indeed satisf\/ied by expression (\ref{89a}).
\end{remark}

\begin{remark}
\label{Remark 5.2} The topological properties of knots and links in $S^3$ and in ${\mathbb R}^3$ are equal. Therefore, expression (\ref{89a}) also describes the Wilson line expectation values for the quantum Chern--Simons theory in ${\mathbb R}^3$ and, in fact, equation~(\ref{89a}) is in agreement with  the results which can  be obtained by means of  standard perturbation theory~\cite{EG}.
\end{remark}

\section[Abelian Chern-Simons theory on $S^1\times S^2$]{Abelian Chern--Simons theory on $\boldsymbol{S^1\times S^2}$}\label{sec6}

\begin{figure}[t]
\centerline{\includegraphics[width=25mm]{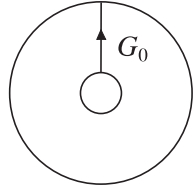}}
\caption{The region of ${\mathbb R}^3$ which is delimited by two spheres $S^2$, one into the other, with their face-to-face points  identif\/ied, provides a description of $S^1\times S^2$. The oriented fundamental loop $G_0 \subset S^1 \times S^2$ is  also represented.}
\label{fig3}
\end{figure}

One can represent $S^1 \times S^2$ by  the region of ${\mathbb R}^3$ which is delimited by two concentric 2-spheres (of dif\/ferent radii), with the convention that  the points on the two surfaces with the same angular coordinates are identif\/ied. The nontrivial knot $G_0$, which can be taken as generator of $H_1(S^1\times S^2, {\mathbb Z}) \simeq {\mathbb Z}$, is shown in Fig.~\ref{fig3}.

\begin{figure}[t]
\centerline{\includegraphics[width=70mm]{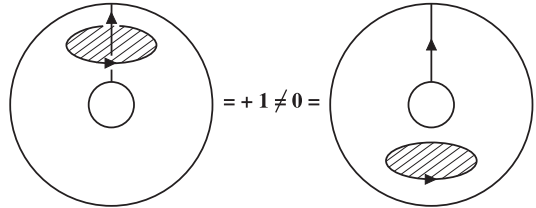}}
\caption{The trivial knot surrounding the non trivial knot $G_0$   is moved down (via an ambient isotopy). The intersection number of its associated surface~-- given by a disc~--   with $G_0$ goes from unity to~0.}
\label{fig4}
\end{figure}

Let us recall that, since $H_2 (S^1 \times S^2, {\mathbb Z})$ is not trivial,  the linking number of two knots may not be well  def\/ined in $S^1 \times S^2$;   one example is shown in Fig.~\ref{fig4}.

Dif\/ferently from $S^3$, the manifold $ S^1 \times S^2$ has nontrivial cohomology and homology groups. While $H_D^3 \left( {S^1\times S^2,{\mathbb Z}} \right)$ is still canonically isomorphic to ${\Omega ^3\left( {S^1\times S^2} \right)}
/ {\Omega _{\mathbb Z}^3 \left( {S^1\times S^2}
\right)}$, the group $H_D^1 \left(
{S^1\times S^2,{\mathbb Z}} \right)$ has the structure of a non trivial af\/f\/ine bundle over
the second integral cohomology group $H^2\left( {S^1 \times S^2 ,{\mathbb Z}} \right)\simeq {\mathbb Z}$. As shown in Fig.~\ref{fig1}, one can then represent $H_D^1 \left( {S^1\times S^2,{\mathbb Z}} \right)$  by means of a collection of f\/ibres over the base space ${\mathbb Z}$, each f\/ibre has a linear space structure and is isomorphic
to ${\Omega ^1\left( {S^1\times
S^2} \right)} / {\Omega _{\mathbb Z}^1 \left( {S^1\times S^2}
\right)}$. For the f\/iber over $0 \in {\mathbb Z}$ one can choose the trivial vanishing connection as canonical origin, so that this f\/ibre can actually be identif\/ied with
${\Omega ^1\left( {S^1\times
S^2} \right)} / {\Omega _{\mathbb Z}^1 \left( {S^1\times S^2}
\right)}$. The f\/iber over $n \in {\mathbb Z}$, with $n \not= 0$, has not a canonical origin, but one can f\/ix  an origin and each element of this f\/ibre will be written as a sum of this origin with an element of ${\Omega ^1\left( {S^1\times
S^2} \right)} / {\Omega _{\mathbb Z}^1 \left( {S^1\times S^2}
\right)}$.

\subsection{Structure of the functional measure}\label{sec6.1}

The choice of an  origin on each f\/ibre of the af\/f\/ine bundle $H_D^1 \left(
{S^1\times S^2,{\mathbb Z}} \right)$  def\/ines of a section $s$ of $H_D^1 \left( {S^1\times S^2,{\mathbb Z}} \right)$ over the discrete base space ${\mathbb Z}\cong H^2\left( {S^1 \times S^2 ,{\mathbb Z}} \right)$, with the convention that $s\left( 0
\right)=\left[ 0 \right]\in H_D^1 \left( {S^1\times S^2,{\mathbb Z}} \right)$.  In agreement with property ({\bf M1}), the quantum measure space ${\cal H}^1_D (S^1 \times S^2, {\mathbb Z})$ can also be understood as an af\/f\/ine bundle over ${\mathbb Z}$, and        the section $s$ will be used to make the structure of the functional integral explicit. Therefore,  one can actually admit distributional values for $s$ and, in fact, it is convenient to def\/ine the section $s $ with values in  ${\widetilde H}_D^1 \left(
{S^1\times S^2,{\mathbb Z}} \right)$.

\begin{definition}
\label{Definition6.1.1} The simplest choice for $s$ is suggested by the additive structure of the base space. More precisely, let us pick up a nontrivial 1-cycle (or oriented knot)  $G_0$ which is directed along the $S^1$ component of $S^1 \times S^2$ and is a generator of $H_1(S^1\times S^2, {\mathbb Z})\simeq {\mathbb Z}$. If  $[\gamma_0] \in  {\widetilde H}_D^1 \left( {S^1\times S^2,{\mathbb Z}} \right)$ denotes the  DB class which is canonically  associated with $G_0$, we shall consider the section
\begin{gather}
 s : \ {\mathbb Z}   \to  {\widetilde H}_D^1 \left( {S^1\times S^2,{\mathbb Z}} \right),\nonumber \\
 \phantom{s :{}} \ n   \mapsto   s\left( n \right) \equiv  n  \left[ {\gamma _0 } \right]   .  \label{92}
\end{gather}

Each element $ [A]$ of $ {\widetilde H}_D^1 \left( {S^1\times S^2,{\mathbb Z}} \right) $ (and of ${\cal H}^1_D (S^1 \times S^2, {\mathbb Z})$) can then be written as
\begin{gather*}
\left[ A \right]=n  \left[ {\gamma _0 } \right]+\left[
\alpha \right]  ,
\end{gather*}
for some integer $n$ and  $\left[ \alpha \right]\in {\Omega ^1\left( {S^1\times S^2} \right)}
/ {\Omega _{\mathbb Z}^1 \left( {S^1\times S^2}
\right)}$;  and the functional measure takes the form
\begin{gather}
 d \mu_k ( [A]) = \sum_{n= - \infty }^{+ \infty}  D [ \alpha ]
 \exp \left \{ 2i\pi k \int_{S^1 \times S^2 }  \left(  n  [ \gamma _0 ] + [ \alpha]  \right) \ast  \left( n [ \gamma _0] + [ \alpha ]  \right)    \right\}   .
\label{94}
\end{gather}
\end{definition}

\begin{remark}
\label{Remark 6.1.1} Because of the translational invariance of the quantum measure,  the particular choice (\ref{92}) of the section $s$ will play no role in the computation of the observables. In fact, a modif\/ication of the origin of each f\/iber of ${\cal H}^1_D (S^1 \times S^2, {\mathbb Z})$ can be achieved by means of an element of ${\Omega ^1\left( {S^1\times S^2} \right)} /
{\Omega _{\mathbb Z}^1 \left( {S^1\times S^2} \right)}$.
\end{remark}

Expression (\ref{94}) can be written as
\begin{gather}
d \mu_k ( [A]) =   \sum_{n= - \infty }^{+ \infty}  D [ \alpha ]
\exp \left \{ 2i\pi k \int_{S^1 \times S^2 }   [ \alpha]  \ast   [ \alpha ]  \right \}
 \exp \left \{ 4i\pi k n \int_{S^1 \times S^2 }    [ \alpha]  \ast   [ \gamma _0]  \right \}
 \nonumber \\
\phantom{d \mu_k ( [A]) =}{} \times  \exp \left \{ 2i\pi k n^2 \int_{S^1 \times S^2 }    [ \gamma_0]  \ast   [ \gamma _0]  \right \}     .
\label{95}
\end{gather}
As usual, in order to def\/ine $[ \gamma_0]  \ast   [ \gamma _0] \in  {\widetilde H}_D^3 \left(
{S^1\times S^2,{\mathbb Z}} \right)$ we shall introduce a framing $G_{0f} $ for the knot $G_0$ and, in agreement with equations (\ref{41}) and  (\ref{42}), we def\/ine  $[ \gamma_0]  \ast   [ \gamma _0] \equiv [ \gamma_0]  \ast   [ \gamma _{0f} ] = [0] \in {\widetilde H}^3_D ( S^1 \times S^2, {\mathbb Z})$.  Therefore, with integers $k$ and $n$, the last factor entering expression (\ref{95}) is well def\/ined and it is equal to the identity. So, one obtains
\begin{gather}
d \mu_k ( [A])  = \sum_{n= - \infty }^{+ \infty}  D [ \alpha ]   \exp \left \{ 2i\pi k \int_{S^1 \times S^2 }   [ \alpha]  \ast   [ \alpha ]  \right \}
 \exp \left \{ 4i\pi k n \int_{S^1 \times S^2 }    [ \alpha]  \ast   [ \gamma _0]  \right \}
   ,
\label{96}
\end{gather}
with $\left[ \alpha \right]\in {\Omega ^1\left( {S^1\times S^2} \right)}
/ {\Omega _{\mathbb Z}^1 \left( {S^1\times S^2}
\right)}$.

\subsection{Zero mode}\label{sec6.2}

\begin{definition}
\label{Definition6.2.2} Let $S_0 $ be a oriented 2-dimensional sphere which is embedded in
$ S^1 \times S^2$ in such a way that it can represent a generator of $H_2 (S^1\times S^2, {\mathbb Z})$.

 $S_0$ is isotopic with the component $S^2$ of  $S^1 \times S^2$ and,  if one represents  $S^1 \times S^2$ by  the region of~${\mathbb R}^3$ which is delimited by two concentric spheres, $S_0 $ can  just be represented by a third concentric sphere. We shall denote by $\beta_0$ the distributional 1-form which is globally def\/ined in $S^1 \times S^2$ and has support on $S_0$; the overall  sign of  $\beta_0 $ is f\/ixed by the orientation of $S_0$ so that
\begin{gather}
\int_{G_0} \beta_0 = 1  .
\label{97}
\end{gather}
Since the boundary of the closed surface $S_0$ is trivial, one has $ d \beta_0 = 0 $.  For any given  real parameter $x $, the 1-form $x \beta_0 $ is also globally def\/ined in $S^1 \times S^2$; let us denote by $[ x \beta_0] \in {\Omega ^1\left( {S^1\times
S^2} \right)} / {\Omega _{\mathbb Z}^1 \left( {S^1\times S^2}
\right)}$ the class  which is represented by the form $x \beta_0 $.
\end{definition}

\begin{proposition}
\label{Proposition6.2.1} For each value $m$ of the integer residues ${\rm mod}\, 2k$,  the Chern--Simons measure~\eqref{96} on $S^1 \times S^2$, with nontrivial coupling constant $k$, satisfies the relation
\begin{gather}
d \mu_k ( [A])  = d \mu_k ( [A] + [ (m/2k) \beta_0 ])   .
\label{98}
\end{gather}
\end{proposition}

\begin{proof} From expression (\ref{96}) one f\/inds
\begin{gather}
d \mu_k ( [A]+ [ (m/2k) \beta_0 ])   \nonumber\\
\qquad{}=
\sum_{n= - \infty }^{+ \infty}  D [ \alpha ]   \exp \left \{ 2i\pi k \int_{S^1 \times S^2 }   [ \alpha]  \ast   [ \alpha ]  \right \}
 \exp \left \{ 4i\pi k n \int_{S^1 \times S^2 }    [ \alpha]  \ast   [ \gamma _0]  \right \}   \nonumber \\
\qquad{}  \times \exp \left \{ 4i\pi k \int_{S^1 \times S^2 }   [ \alpha]  \ast   [ (m/2k) \beta_0 ]  \right \}     \exp \left \{ 2i\pi k \int_{S^1 \times S^2 }   [ (m/2k) \beta_0]  \ast   [ (m/2k) \beta_0 ]  \right \}   \nonumber \\
\qquad{} \times  \exp \left \{ 4i\pi k n \int_{S^1 \times S^2 }   [ (m/2k) \gamma_0]  \ast   [ \eta_0 ]  \right \}  ,
\label{99}
\end{gather}
where the integer $m$ takes the values $m=0,1,2,\dots, 2k-1$. From the equality $ d \beta_0 = 0 $ it  follows that
\begin{gather*}
 4i\pi k \int_{S^1 \times S^2 }   [ \alpha]  \ast   [ (m/2k) \beta_0 ] = 2 i \pi m
 \int_{S^1 \times S^2 }    \alpha \wedge  d \beta_0  = 0  ,
\end{gather*}
where $\alpha \in \Omega ^1\left( {S^1\times S^2} \right)$ represents the class $[\alpha ]$,
\begin{gather*}
2i\pi k \int_{S^1 \times S^2 }   [ (m/2k) \beta_0]  \ast   [ (m/2k) \beta_0 ] = i \pi ( m^2 / 2k)  \int_{S^1 \times S^2 }    \beta_0 \wedge d  \beta_0  = 0  .
\end{gather*}
Finally, relation (\ref{97}) implies
\begin{gather*}
\exp \left \{ 4i\pi k n \int_{S^1 \times S^2 }   [ (m/2k) \beta_0]  \ast   [ \gamma_0 ]  \right \} = \exp \left \{ 2i\pi  n m  \int_{G_0 }   \beta_0  \right \}  = 1  .
\end{gather*}
Therefore expressions (\ref{99}) and (\ref{96}) are equal.
\end{proof}

\subsection{Values of the observables}\label{sec6.3}

Let us consider an oriented coloured and framed   link $L $ in $  S^1 \times S^2$; without loss of generality, one can always assume that $L$ does not intersect the knot $G_0$. In agreement with equation (\ref{97}), the integral
\begin{gather*}
N_0(L) = \int_{L}  \beta_0
\end{gather*}
takes integer values; more precisely, $N_0(L)$ is equal to the sum of the  intersection numbers (weighted with the charges of the link components)  of the link $L$ with the surface $S_0$.

\begin{theorem}\label{Theorem2}
Given a link $L \subset  S^1 \times S^2$,
\begin{itemize}
\item when $N_0(L) \not\equiv 0 \mod 2k$,  one finds  $\left\langle W(L) \right\rangle _{k} =0 $;
\item  whereas for $N_0(L) \equiv 0 \mod 2k$, one has
\begin{gather}
\left\langle W(L) \right\rangle _{k}  =   \exp \left \{ -( 2 i \pi / 4k)   \int_{S^1 \times S^2} \eta_L  \wedge d \eta_L \right  \}    ,
\label{113}
\end{gather}
where $\eta_L  \wedge d  \eta_L$ is defined by means of the framing procedure.
\end{itemize}
\end{theorem}

\begin{proof}  The expectation value of the Wilson line operator is given by
\begin{gather}
\left\langle W(L)
\right\rangle _{k}  =   Z_k^{-1} \int d\mu _k  (
[ A ] )   \exp \left \{ {2 i \pi  \int_{S^1 \times S^2} [ A ] *[ \eta_L ]} \right  \}   ,
\label{103}
\end{gather}
where $d\mu _k  ( [ A ] ) $ is shown in equation (\ref{96}) and
\begin{gather*}
Z_k =  \int d\mu _k  ( [ A ] )  .
\end{gather*}
Equation (\ref{98}) implies that $W(L)$ satisf\/ies the following relation
\begin{gather}
\left\langle W(L)
\right\rangle _{k}   =    Z_k^{-1}\frac{1}{2k} \sum_{m=0}^{2k-1} \int d\mu _k  (
[ A ] + [(m/2k)\beta_0] )   e^{  2 i \pi  \int_{S^1 \times S^2} ([ A] + [(m/2k)\beta_0] ) *[ \eta_L ]   }  \nonumber \\
\phantom{\left\langle W(L) \right\rangle _{k}}{}  = Z_k^{-1}
\int d\mu _k  (
[ A ])   e^{  2 i \pi  \int_{S^1 \times S^2} [ A] *[ \eta_L ]   }   \frac{1}{2k} \sum_{m=0}^{2k-1} e^{  2 i \pi  \int_{S^1 \times S^2} [ (m/2k)\beta_0]  *[ \eta_L ]   }\nonumber \\
\phantom{\left\langle W(L) \right\rangle _{k}}{} = \left\langle W(L)
\right\rangle _{k}  \frac{1}{2k} \sum_{m=0}^{2k-1} \exp \left \{  2 i \pi (m /2k)  \int_{L}  \beta_0  \right \}  .
\label{105}
\end{gather}
 One has
\begin{gather*}
\frac{1}
{{2k}}\sum\limits_{m = 1}^{2k - 1} {\exp \left\{ {{{2i\pi N_0 \left( L \right)m} / {2k}}} \right\}}  = \left\{ \begin{array}{ll}
  1 & \text{if} \quad N_0 \left( L \right) \equiv 0 \mod 2k,  \\
  0 & \text{otherwise}.
\end{array}  \right.
 \end{gather*}
Therefore equation (\ref{105}) shows that, when $N_0(L) \not\equiv 0 \mod 2k$,  the expectation value $\left\langle W(L) \right\rangle _{k} $ is vanishing.

Let us now consider the case in which $N_0(L) \equiv 0 $ mod $2k$. Because of Proposition~\ref{Proposition4.4.1}, we only need to discuss the case  $N_0(L) =0 $. In fact, if $N_0(L) = 2k p $ for some integer  $p \not= 0$,  at least one of the link components $C\subset L$  intersects $S_0$; one can then modify the value $q_C$ of its charge according to $q_C \rightarrow q_C - 2k p$ so that $N_0(L)$ vanishes.  According to the decomposition $[A] = n [\gamma_0 ] + [\alpha ]$, one f\/inds
\begin{gather*}
\exp \left \{ 2 i \pi  \int_{S^1 \times S^2} [ A ] *[ \eta_L ] \right \}    =
\exp \left \{ 2 i \pi n \int_{S^1 \times S^2} [ \gamma_0 ] *[ \eta_L ] \right \}  \exp \left \{ 2 i \pi  \int_{S^1 \times S^2} [ \alpha ] *[ \eta_L ] \right \} \nonumber \\
 \phantom{\exp \left \{ 2 i \pi  \int_{S^1 \times S^2} [ A ] *[ \eta_L ] \right \} }{}  = \exp \left \{ 2 i \pi  \int_{S^1 \times S^2} [ \alpha ] *[ \eta_L ] \right \}  ,
\end{gather*}
where the last equality is a consequence of  the identity $ [ \gamma_0 ] *[ \eta_L ] = [0] \in  {\widetilde H}_D^3 \left( {S^1\times S^2,{\mathbb Z}} \right)$, which follows from the framing procedure.  Then, from equation (\ref{103}) one gets
\begin{gather}
\left\langle W(L)
\right\rangle _{k}  =   Z_k^{-1} \int \sum_{n= - \infty }^{+ \infty}  D [ \alpha ]    e^{ 2i\pi k \int_{S^1 \times S^2 }   [ \alpha]  \ast   [ \alpha ] }
e^{ 4i\pi k n \int_{S^1 \times S^2 }    [ \alpha]  \ast   [ \gamma _0] }
  e^{ 2 i \pi  \int_{S^1 \times S^2} [ \alpha ] *[ \eta_L ] }   .
\label{109}
\end{gather}
When $N_0(L) =0 $, the link $L$ is homological trivial and one can f\/ind a Seifert  surface for $L$. More precisely, in agreement with Proposition~\ref{Proposition4.6.1} and equation (\ref{78}), one can substitute $L$ with its simplicial satellite ${\overline L}$, def\/ined in Section~\ref{sec4}, whose components have unitary charges. The oriented framed link ${\overline L}\subset S^1 \times S^2$ also is homologically trivial and it is the boundary of an oriented surface that we shall denote by $\Sigma_{\overline L} \subset S^1 \times S^2$.   Let  $\eta_L $ be the distributional 1-form with support on $\Sigma_{\overline L}$ which is globally def\/ined in $S^1 \times S^2$;  because of Proposition~\ref{Proposition4.6.1},  in the Chech--de~Rham description of the DB classes, $[ \eta_L ] $ can then be represented by the sequence  $( \eta_L , 0 , 0 )$.  The 1-form $ (1/2k) \eta_L$ also is globally def\/ined in  $S^1 \times S^2$ and we shall denote by $[ (1/2k)  \eta_L] $ the DB class which is represented by the form $ (1/2k) \eta_L$. By construction,
\begin{gather}
\exp \left \{ - 4i \pi k   \int_{S^1 \times S^2} [\alpha ] \ast [ (1/2k)  \eta_L]  \right \} =
\exp \left \{ - 2i \pi    \int_{S^1 \times S^2} [\alpha ] \ast [   \eta_L]  \right \}  ,
\label{110}
\end{gather}
and the condition $N_0(L) =0 $ (or $N_0(L) \equiv 0 $ mod $2k$) implies  that, for integer $n$,
\begin{gather}
\exp \left \{ - 4i \pi k n  \int_{S^1 \times S^2} [ (1/2k)  \eta_L] \ast [\gamma_0] \right \} = 1  .
\label{111}
\end{gather}
By means of the substitution $[\alpha ] \rightarrow [\alpha] - [ (1/2k) \eta_L] $ and with the help of equations~(\ref{110}) and~(\ref{111}), expression~(\ref{109}) assumes the form
\begin{gather*}
\left\langle W(L) \right\rangle _{k}  =   \exp \left \{ -( 2 i \pi / 4k)   \int_{S^1 \times S^2} \eta_L  \wedge d  \eta_L \right  \}  Z_k^{-1}  Z_k   .
\end{gather*}
Therefore, assuming  $Z_k \not= 0 $, when $N_0(L) \equiv 0 $~mod $2k$  one gets
\begin{gather*}
\left\langle W(L) \right\rangle _{k}  =   \exp \left \{ -( 2 i \pi / 4k)   \int_{S^1 \times S^2} \eta_L  \wedge d  \eta_L \right  \}    ,
\end{gather*}
and this concludes the proof.
\end{proof}

\begin{figure}[t]
\centerline{\includegraphics[width=70mm]{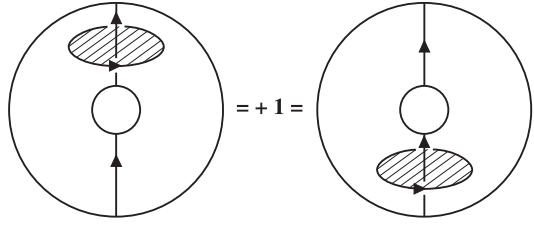}}
\caption{An example of conservation of the intersection number under ambient isotopy  for a globally trivial 1-cycle.}
\label{fig5}
\end{figure}

\begin{remark}
 \label{Remark6.3.1} Expression (\ref{113}) formally coincides with the result (\ref{89}) which has been obtained in the case $M_3 \sim S^3$. It should be noted that the integral  (which appears in equation~(\ref{113}))
\begin{gather}
\int_{S^1 \times S^2} \eta_L  \wedge d  \eta_L \equiv \int_{S^1 \times S^2} \eta_L  \wedge d  \eta_{ {\overline L}_f} = \int_{{\overline L}_f} \beta_L  ,
\label{114}
\end{gather}
where ${\overline L}_f$ denotes the framing of $\overline L$, is well def\/ined  because it does not depend on the choice of the Seifert surface of ${\overline L}$.   Indeed suppose that, instead of $\Sigma_{\overline L} $, we take $\Sigma^{ \prime}_{\overline L} $ as Seifert surface for the link $\overline L$.
The dif\/ference between the intersection number (\ref{114}) of ${\overline L}_f$ with $\Sigma^{ \prime}_{\overline L} $ and $\Sigma_{\overline L} $  is given by the intersection number of ${\overline L}_f$ with the closed surface $\Sigma^{ \prime}_{\overline L} \cup \Sigma^{-1}_{\overline L} $. This surface could be nontrivial in $S^1 \times S^2$ but, since $\overline L$ is homologically trivial, ${\overline L}_f$ also is homologically trivial and then its intersection number with a closed surface vanishes. The example of Fig.~\ref{fig5} illustrates the ambient isotopy invariance  of the intersection number of a homologically trivial link with the Seifert surface of a trivial knot  in $S^1 \times S^2$.
\end{remark}

\section[Abelian Chern-Simons theory on $S^1\times \Sigma _g $]{Abelian Chern--Simons theory on $\boldsymbol{S^1\times \Sigma _g }$}\label{sec7}

Let us now consider the manifold  $M_3 \sim S^1\times \Sigma _g $ where
$\Sigma _g $ is a closed Riemann surface  of genus $g\ge 1$. In this case, the computation of the Chern--Simons observables is rather similar to the computation when $M_3 \sim S^1 \times S^2$.   So, we shall brief\/ly illustrate the main steps of the construction.

As it has been mentioned in Section~\ref{sec1}, $H_D^1 ( S^1 \times \Sigma_g , {\mathbb Z})$ has the structure of a af\/f\/ine bundle over $H^2(S^1 \times \Sigma_g , {\mathbb Z}) \sim {\mathbb Z}^{2g+1} $ with  $\Omega^1( S^1 \times \Sigma_g) / \Omega^1_{\mathbb Z} ( S^1 \times \Sigma_g)$ acting canonically on each f\/ibre by translation. In agreement with property~({\bf M1}), the functional space ${\cal H}_D^1 ( S^1 \times \Sigma_g , {\mathbb Z})$ is assumed to have  the same structure of $H_D^1 ( S^1 \times \Sigma_g , {\mathbb Z})$ and, in order to f\/ix a origin in each f\/ibre, we need to introduce a section $s : {\mathbb Z}^{2g+1} \rightarrow {\cal H}_D^1 ( S^1 \times \Sigma_g , {\mathbb Z})$.

\begin{definition}
\label{Definition7.1} Let the nonintersecting  oriented framed knots $ \{ G_0 , G_1, \dots, G_{2g} \} $ in $S^1\times \Sigma _g $ represent the generators of $H_1 \left( {S^1\times \Sigma _g ,{\mathbb Z}} \right)$.  For each  $j=0,1,\dots, 2g$,
we shall denote by $[ \gamma_j] \in {\widetilde H}_D^1 ( S^1 \times \Sigma_g , {\mathbb Z})$ the DB class which is canonically associated with the knot $G_j$.
\end{definition}

\begin{definition}
\label{Definition7.2} If the  elements of ${\mathbb Z}^{2g+1} $ are represented by vectors
\begin{gather*}
\vec n \equiv \left (n_0 ,  n_1 , n_2 , \dots , n_{2g} \right ) \in   {\mathbb Z}^{2g+1}
  ,
\end{gather*}
 a possible choice for the section $s$ is given by
\begin{gather*}
 s : \ {\mathbb Z}^{2g+1}  \to  {\widetilde H}_D^1 \left( {S^1\times \Sigma_g ,{\mathbb Z}} \right),\nonumber \\
\phantom{s : {}} \ \vec  n   \mapsto   s\left( \vec n \right) =   [  n   \gamma  ] \equiv  \vec n \cdot [\vec \gamma  ] =  \sum_{j =0}^{2g} n_j [\gamma_j ]   . 
\end{gather*}

Each class $[A] \in  {\widetilde H}_D^1 ( S^1 \times \Sigma_g , {\mathbb Z})$ can then be written as
\begin{gather*}
[A ] = [  n  \gamma  ] + [\alpha ]  ,
\end{gather*}
for certain $\vec n $ and $[\alpha ] \in \Omega^1( S^1 \times \Sigma_g) / \Omega^1_{\mathbb Z} ( S^1 \times \Sigma_g)$. Consequently, the Chern--Simons  functional measure takes the form
\begin{gather}
d \mu_k ( [A])  = \sum_{\vec n}  D [ \alpha ]   \exp \left \{ 2i\pi k \int_{S^1 \times S^2 }   [ \alpha]  \ast   [ \alpha ]  \right \}
 \exp \left \{ 4i\pi k  \int_{S^1 \times S^2 }    [ \alpha]  \ast   [  n  \gamma  ]  \right \}
   ,
\label{117}
\end{gather}
which is the analogue of equation (\ref{96}). The  condition $  [  n  \gamma  ] \ast  [ n  \gamma  ] = 0 \in {\widetilde H}_D^3 ( S^1 \times \Sigma_g , {\mathbb Z})$, which results from the framing procedure, has already been used to  simplify the expression of $d \mu_k ( [A]) $.
\end{definition}

\begin{definition}
\label{Definition7.3}  Let  the oriented closed surfaces $S_j \subset S^1 \times \Sigma_g $, with  $j = 0, 1,\dots, 2g $, represent  the generators of $H_2 ( S^1 \times \Sigma_g, {\mathbb Z} )\sim {\mathbb Z}^{2g+1}$.  We shall denote by  $\beta_j \in  {\widetilde H}_D^1 \left( {S^1\times \Sigma_g ,{\mathbb Z}} \right)$ the distributional 1-form which is globally def\/ined in $S^1 \times \Sigma_g $ and has support on $S_j$. One can choose the generators of $H_2 ( S^1 \times \Sigma_g, {\mathbb Z} )$  in such a way that  the following orthogonality relations are satisf\/ied
\begin{gather*}
\int_{G_i} \beta_j = \delta_{ij}  , \qquad i,j = 0,1,\dots, 2g  .
\end{gather*}

Since $S_j$ are closed surfaces, one has $d \beta_j = 0 $. For any real parameter $x$, the 1-form $ x \beta_j$ also is globally def\/ined in $S^1 \times \Sigma_g $ and the corresponding class, which can be represented by $x \beta_j$, will be denoted by $[x \beta_j ] \in \Omega^1( S^1 \times \Sigma_g) / \Omega^1_{\mathbb Z} ( S^1 \times \Sigma_g)$.  The arguments that have been presented to prove Proposition~\ref{Proposition6.2.1} can also be used to prove the following
\end{definition}

\begin{proposition}
\label{Proposition7.1} The quantum measure \eqref{117} of the Chern--Simons theory on $S^1 \times \Sigma_g$, with nontrivial coupling constant $k$, satisfies the relation
\begin{gather*}
d \mu_k ( [A])  = d \mu_k ( [A] + [ (m/2k) \beta_j ])   .
\end{gather*}
for  $m=0,1,2,\dots, 2k-1$ and for each value of $j =0,1,\dots, 2g$.
\end{proposition}

 Finally, the expectation values of the Wilson line operators are determined by the following

\begin{theorem}
\label{Theorem3}  Let $L$ be a oriented coloured framed link in $ S^1 \times \Sigma_g$. For  each $j =0,1,\dots, 2g$, let us introduce the integer
\begin{gather*}
N_j(L) = \int_{L}  \beta_j  .
\end{gather*}
 Then
\begin{itemize}\itemsep=0pt
\item  when  $N_j(L) \not\equiv 0\mod 2k$ for at least one value of $j=0,1,\dots, 2g$,  one has   $\left\langle W(L) \right\rangle _{k} =0 $ ;
\item whereas when $N_j(L) \equiv 0 \mod 2k$ for all values of  $j=0,1,\dots, 2g$, one finds
\begin{gather}
\left\langle W(L) \right\rangle _{k}  =   \exp \left \{ -( 2 i \pi / 4k)   \int_{S^1 \times \Sigma_g} \eta_L  \wedge d  \eta_L \right  \}    ,
\label{121}
\end{gather}
 where $\eta_L  \wedge d  \eta_L$ is defined by means of the framing procedure.
\end{itemize}
\end{theorem}

\begin{proof}  The proof is similar to  the proof  of Theorem~\ref{Theorem2}. In fact,
when  $N_j(L) \not\equiv 0$~mod~$2k$ for at least one value of $j=0,1,\dots, 2g$,   Proposition~\ref{Proposition7.1} implies that the Chern--Simons expectation value $ \left\langle W(L) \right\rangle _{k} $ vanishes.  On the other hand, when $N_j(L) \equiv 0$~mod~$2k$ for all values of  $j=0,1,\dots, 2g$, the substitution $[\alpha ] \rightarrow [\alpha] - [ (1/2k) \eta_L] $ in the functional measure~(\ref{117}) leads to the equation~(\ref{121}). It should be noted that expression  (\ref{121}) is well def\/ined  because the link $L$ and then its framing~$L_f$ are homologically trivial. \end{proof}

\section{Surgery rules}\label{sec8}

For the quantum Abelian Chern--Simons theory on the manifolds $S^1 \times S^2$ and $S^1 \times \Sigma_g $ (and, in general, in any nontrivial 3-manifold), the standard gauge theory approach which is based on the gauge group $U(1)$ is in principle well def\/ined but presents some technical dif\/f\/iculties, which are related, for instance,  to the implementation of the gauge f\/ixing procedure and the determination of the Feynman propagator. As a matter of facts, by means of the usual methods of quantum gauge theories, the  computation of the Chern--Simons observables  in a nontrivial 3-manifold has never been explicitly produced.

In order to determine the  Wilson line expectation values in $M_3 \not\sim S^3$, one can use for instance the surgery rules of the Reshetikhin--Turaev  type \cite{RT} as  developed by Lickorish~\cite{LIC} and by Morton and Strickland~\cite{MOST}.
In this section, we outline the surgery method which turns out to produce the Chern--Simons observables  for the manifolds $S^1 \times S^2$ and $S^1 \times \Sigma_g $ in complete agreement with the results obtained in the DB approach of the path-integral.

Every closed orientable connected 3-manifold $M_3$ can be obtained by Dehn  surgery on $S^3$ and admits a surgery presentation~\cite{ROL} which is described by a framed surgery link ${\cal L}\subset S^3$ with integer surgery coef\/f\/icients.  Each surgery coef\/f\/icient specif\/ies the framing of the corresponding component of ${\cal L}$ because it coincides with the linking number of this component  with its framing. The manifold $S^1 \times S^2$ admits a presentation with surgery link given by the unknot with vanishing surgery coef\/f\/icient, whereas $S^1 \times S^1 \times S^1$ for example corresponds to the Borromean rings with vanishing surgery coef\/f\/icients.  Any oriented coloured framed link $L \subset M_3$ can be described by a link $L^{ \prime } = L \cup {\cal L} $ in $S^3$ in which:

\begin{itemize}\itemsep=0pt

\item the surgery link  ${\cal L}$ describes the surgery instructions corresponding to a presentation of $M_3$ in terms of Dehn surgery on $S^3$;

\item  the remaining components of $L^{ \prime } $ describe how $L$ is placed in $M_3$.

\end{itemize}

Assuming that the expectation values of the Wilson line operators form a complete set of observables, one can f\/ind \cite{EG} consistent surgery rules, according to which the expectation value of the Wilson line operator $W(L)$ in $M_3$ can be written as a ratio
\begin{gather}
\langle W ( L) \rangle_k   \vert_{M_3} =   \langle W ( L)  W({\cal L}) \rangle_k   \vert_{S^3}  \;  / \;   \langle  W({\cal L}) \rangle_k  \vert_{S^3}  ,
\label{122}
\end{gather}
where to each component of the surgery link is associated a  particular colour state $\psi_0$. Remember that, for f\/ixed integer $k$, the colour space coincides with  space of residue classes of integers mod~$2k$, which has a canonical ring structure; let $\chi_j$ denote the residue class associated with the integer $j$. Then, the colour state $\psi_0 $ is given by
\begin{gather*}
\psi_0 = \sum_{j=0}^{2k -1}  \chi_j  .
\end{gather*}
One can verify that the surgery rule (\ref{122}) is well def\/ined and consistent; in fact,  expression~(\ref{122}) is invariant under Kirby moves~\cite{KI}. Finally, one can check that, according to the surgery formula~(\ref{122}), the expectation values of the Wilson line operators in $S^1 \times S^2$ and in $S^1 \times \Sigma_g$ are given precisely by the expressions of Theorems~\ref{Theorem2} and \ref{Theorem3}, which have been  obtained by means of the DB cohomology.

\section{Conclusions}\label{sec9}

In the standard f\/ield theory formulation of Abelian gauge theories, the (classical f\/ields)  conf\/igu\-ra\-tion space is taken to be  the set of 1-forms modulo closed forms. But  when  the observables of the theory are given by the exponential of the holonomies which are associated with oriented loops,  the classical conf\/iguration space is actually given by the set of 1-forms modulo forms of integer periods;  that is, the classical conf\/iguration space indeed coincides with space of the Deligne--Beilinson cohomology classes. So, in this article we have considered the Abelian Chern--Simons gauge theory, in which a complete set of observables is given by  the set of exponentials of the holonomies which are associated with oriented knots or links in a 3-manifold $M_3$. We have explored the main properties of the quantum theory and of the corresponding quantum functional integral, which enters the computation of the observables, when the path-integral is really  def\/ined over the Deligne--Beilinson classes. Within this new approach, we have produced an explicit path-integral computation of the Chern--Simons link invariants in a class of torsion-free 3-manifolds. In facts, we have not used any standard gauge-f\/ixing and perturbative method, as it has been done so far in literature. Our results are based on an explicit  non-perturbative path-integral computation and are exact results.

Let us brief\/ly summarize the main issues of our article.
In Sections~\ref{sec2} and~\ref{sec3} we have discussed a few technical points which are important for the computation of the observables. The basic def\/initions and properties of the DB cohomology together with a distributional extension of the space of the equivalence classes have been illustrated. Then we have shown how the framing procedure, which is used to give a topological meaning to the self-linking number, can be naturally def\/ined also in the DB context.  The general features of the Abelian Chern--Simons theory  in a generic 3-manifold $M_3$ have been derived in Section~\ref{sec4}.  The main achievements concerning the observables  are the ``colour periodicity'' property (Proposition~\ref{Proposition4.4.1}), the ``ambient isotopy inva\-riance'' (Proposition~\ref{Proposition4.5.1}) and the validity of appropriate ``satellite relations'' (Proposition~\ref{Proposition4.6.1}).  With respect to the standard f\/ield theory approach, our proofs extend the validity of these properties from ${\mathbb R}^3$ to a generic (closed and oriented) manifold  $M_3$.

The Abelian Chern--Simons theory formulated in $S^3$ is discussed in Section~\ref{sec5} and its solution is given by Theorem~\ref{Theorem1}; in this case, the outcome is in agreement with the results obtained by means of standard perturbation theory in ${\mathbb R}^3$.  The expressions of the observables for the Chern--Simons theory formulated in $S^1 \times S^2 $ and in a generic 3-manifold of the type $S^1 \times \Sigma_g$ are contained in Theorems~\ref{Theorem2} and~\ref{Theorem3};  in the standard f\/ield theory approach, no proof of these theorems actually exists.

Finally, we have checked the validity our path-integral results by means of an alternative ``combinatorial method''. Indeed, the link invariants def\/ined in the Chern--Simons theory  are related to the link invariants def\/ined by means of the quantum group methods of  Reshetikhin and Turaev.  Given a surgery presentation in $S^3$ of a generic 3-manifold $M_3$  and knowing the values of the link invariants in $S^3$, one can use the surgery method of Lickorish and  Morton--Strickland to determine the values of the link invariants in  $M_3$. As far as the Abelian Chern--Simons is concerned, we have presented the basic aspects of this surgery method  in Section~\ref{sec8}. We have verif\/ied that the expression of the link invariants  for the manifolds   $S^1 \times S^2 $ and  $S^1 \times \Sigma_g$, which are described by Theorems~\ref{Theorem2} and~\ref{Theorem3},  precisely coincide with the results obtained by means of the surgery method.

Clearly, in the case of a generic 3-manifold,  the general features of the Deligne--Beilinson approach to the Abelian Chern--Simons functional integral remain to be fully explored. Possible applications of this formalism to the non-Abelian Chern--Simons theory would also give new hints on the topological meaning of the polynomial link invariants. Finally, we mention that extensions of Deligne--Beilinson cohomology approach to the topological f\/ield theories  in lower dimensions can easily be produced, but the resulting structure of the observables appears to be quite elementary. Presumably, applications in higher dimensions will produce more interesting invariants.

\subsection*{Acknowledgements}
We wish to thank Raymond Stora for useful discussions.

\pdfbookmark[1]{References}{ref}
\LastPageEnding

\end{document}